\documentclass[12pt]{JHEP3}
\usepackage{epsfig}
\usepackage{amsmath}
\def\beqra{\begin{eqnarray}} \def\eeqra{\end{eqnarray}}
\def\beqast{\begin{eqnarray*}} \def\eeqast{\end{eqnarray*}}
\def\beq{\begin{equation}}      \def\eeq{\end{equation}}
\def\be{\begin{enumerate}}   \def\ee{\end{enumerate}}

\def\vphi{\varphi}

\def\om{\omega}
\def\Om{\Omega}

\def\pa{\partial}

\newcommand{\pv}[1]{{-  \hspace {-4.0mm} #1}}

\def\raisenot{\raise .5mm\hbox{/}}
\def\nota{\ \hbox{{$a$}\kern-.49em\hbox{/}}}
\def\notA{\hbox{{$A$}\kern-.54em\hbox{\raisenot}}}
\def\notb{\ \hbox{{$b$}\kern-.47em\hbox{/}}}
\def\notB{\ \hbox{{$B$}\kern-.60em\hbox{\raisenot}}}
\def\notc{\ \hbox{{$c$}\kern-.45em\hbox{/}}}
\def\notd{\ \hbox{{$d$}\kern-.53em\hbox{/}}}
\def\notbd{\ \hbox{{$D$}\kern-.61em\hbox{\raisenot}}} 
\def\note{\ \hbox{{$e$}\kern-.47em\hbox{/}}}
\def\notk{\ \hbox{{$k$}\kern-.51em\hbox{/}}}
\def\notp{\ \hbox{{$p$}\kern-.43em\hbox{/}}}
\def\notq{\ \hbox{{$q$}\kern-.47em\hbox{/}}}
\def\notW{\ \hbox{{$W$}\kern-.75em\hbox{\raisenot}}}
\def\notz{\ \hbox{{$Z$}\kern-.61em\hbox{\raisenot}}}
\def\notpa{\hbox{{$\partial$}\kern-.54em\hbox{\raisenot}}}
\def\fo{\hbox{{1}\kern-.25em\hbox{l}}}  

\def\dgg{\dagger}

\def\pix{\pi(x)}

\def\pax{\pa_x}

\def\sign{{\rm sign}\,}

\title{
Fluctuations around Periodic BPS-Density Waves in the Calogero Model
}
\author{
V. Bardek$^{a 1}$, 
J. Feinberg$^{b,c 2}$
and 
S. Meljanac$^{a 3}$\\
${}^a$ {\it Rudjer Bo\v{s}kovi\'c Institute, Bijeni\v cka  c.54, HR-10002 Zagreb,
Croatia }\\
${}^b$ {\it Department of Physics, University of Haifa at Oranim, Tivon 36006,
Israel}\\
${}^c$ {\it Department of Physics, Technion-Israel Inst. of Technology,
Haifa 32000, Israel}\\
${}^1$ E-mail: \email{bardek@irb.hr}\\
${}^2$ E-mail: \email{joshua@physics.technion.ac.il}\\
${}^3$ E-mail: \email{meljanac@irb.hr}\\
}

\abstract
{
The collective field formulation of the Calogero model supports periodic density waves. An important set of such density waves is a two-parameter
family of  BPS solutions of the equations of motion of the collective field theory. One of these parameters is essentially the average particle density,
which determines the period, while the other parameter determines the amplitude. These BPS solutions are sometimes referred to as ``small amplitude
waves" since they undulate around their mean density, but never vanish. We present complete analysis of quadratic fluctuations around these BPS solutions. 
The corresponding fluctuation hamiltonian (i.e., the stability operator) is diagonalized in terms of bosonic creation and annihilation operators which correspond to 
the complete orthogonal set of Bloch-Floquet eigenstates of a related periodic Schr\"odinger hamiltonian, which we derive explicitly. 
Remarkably, the fluctuation spectrum is independent of the parameter which determines the density wave's amplitude. As a consequence, the sum over zero-point 
energies of the field-theoretic fluctuation hamiltonian, and its ensuing normal-ordering and regularization, are the same as in the case of fluctuations around constant 
density background, namely, the ground state. Thus, quadratic fluctuations do not shift the energy density tied with the BPS-density waves studied here, compared to its 
ground state value. Finally, we also make some brief remarks concerning fluctuations around non-BPS density waves. }

\keywords{Calogero Model, Collective-Field Theory, BPS, Solitons, Fluctuations, Floquet-Bloch}
\preprint{}
\begin{document}

\section{Introduction}
\label{section1}
In the limit of infinite number of particles and finite density, the Calogero-Sutherland models \cite{Calogero:1969xj, Sutherland:1971ep} may be reformulated 
in terms of hydrodynamic variables, namely,  the particle density and current. The hamiltonian of this collective field reformulation is given by \cite{Andric:1982jk} 
\beq\label{Hcollective} H_{coll} = \frac{1}{2m} \int dx\, \pax\pi(x)\, \rho(x)\, \pax\pi(x) + \frac{1}{2 m}
\int dx\, \rho(x) \left( \frac{\lambda - 1}{2} \frac{\partial_{x}\rho}{\rho} + \lambda \pv \int \frac{ dy \rho(y)}{x - y}
\right)^{2}  + H_{sing}\,, \eeq 
where  $ \; H_{sing} \; $ denotes a singular contribution \cite{Andric:1988vt, ABJ:1995}
\beq\label{Hsing} H_{sing} =  - \frac{\lambda}{2 m}\,\int dx\,
\rho(x)\,\partial_{x}\left. \frac{P}{x - y} \right|_{y = x} -
\frac{\lambda - 1}{4 m}\,\int dx\, {\partial_{x}}^{2} \left.
\delta(x - y) \right|_{y = x}\,, \eeq 
and $ \; P \;$ is the principal part symbol.

Here, \beq\label{collective} \rho(x) = \sum_{i = 1}^{N} \delta( x
- x_{i}) \eeq is the collective  - or density - field, and
\beq\label{momenta} \pi(x) = - i \frac{\delta}{\delta \rho(x)}
\end{equation}
is its canonically conjugate momentum. It follows from
(\ref{collective}) that the collective field is a positive
operator \beq\label{positiverho} \rho(x) \geq 0\,,\eeq and that it
obeys the normalization condition \beq\label{conservation}
\int\limits_{-\infty}^\infty\,dx \,\rho (x) = N\,. \eeq The latter
constraint is implemented by adding to (\ref{Hcollective}) a term
$\mu\left(\int\limits_{-\infty}^\infty\,dx \,\rho (x) - N\right)$,
where $\mu$ is a Lagrange multiplier (the chemical potential). 

The first term in (\ref{Hsing}) is linear in $\rho(x)$. Therefore, its singular coefficient
$-{\lambda\over 2m}\partial_{x}\left. \frac{P}{x - y} \right|_{y = x} $ amounts to a shift of the 
chemical potential $\mu$ by an infinite constant. The last term in (\ref{Hsing}) is, of course, 
a field independent constant - an infinite shift of energy.

For the sake of being self-contained, we briefly explain in the Appendix how to derive the collective hamiltonian (\ref{Hcollective}) and its
singular part (\ref{Hsing}) from the microscopic Calogero hamiltonian.

It is worth mentioning at this point that the Calogero model
enjoys a strong-weak-coupling duality symmetry
\cite{Minahan:1994ce, halzirn}. At the level of the collective
Hamiltonian (\ref{Hcollective}), these duality transformations
read \beq\label{duality} \tilde\lambda = {1\over\lambda}\,,\quad
\tilde m = -{m\over\lambda}\,,\quad \tilde \mu =
-{\mu\over\lambda}\,;\quad \tilde\rho(x) = -\lambda\rho(x)\,,\quad
{\rm and}\quad \tilde\pi(x) = -{\pi(x)\over\lambda}\,,\eeq and it
is straightforward to see that these transformations leave
(\ref{Hcollective}) (including the chemical potential term)
invariant. The minus signs which occur in (\ref{duality}) are all
important: We interpret all negative masses and
densities as those pertaining to holes, or antiparticles. Thus,
the duality transformations (\ref{duality}) exchange particles and
antiparticles. With this interpretation we always have 
\beq\label{positivity}
{\rho(x)\over m} >0\,,
\eeq
and thus the first two terms in (\ref{Hcollective}) are manifestly positive. 
(For more details see e.g. Section 3 of \cite{BFM:2007}, and references therein.)

One of the important and interesting features of this collective-field formulation of the Calogero model is that it bears soliton solutions, which are known 
explicitly \cite{Jevicki:1991yi, Polychronakos:1994xg, Andric:1994nc, Sen:1997qt}. They arise (to leading order in the ${1\over N}$ expansion) as solutions
of the classical equations of motion resulting from (\ref{Hcollective}) \cite{Jevicki:1979mb}.

Recently there has been considerable renewed interest in solitons in the Calogero-Sutherland models. In \cite{abanov} it was shown that the collective-field theory 
of Calogero model is equivalent to  a quantum version of an integrable Benjamin-Ono equation \cite{BO}. In this way, it can be seen that  the semiclassical one-soliton 
solutions studied in \cite{Jevicki:1991yi, Polychronakos:1994xg, Andric:1994nc, Sen:1997qt}  correspond to a single pole in a certain pole ansatz. 

The authors of  \cite{AJJ:2005, AJJ:2006} have studied a specific duality-based generalization of the hermitian matrix model which is equivalent to a two-family 
Calogero model \cite{Sen:1995zj, Meljanac:2003jj, Bardek:2005cs, BMEPL:2005}.  The multi-vortex solutions of the coupled BPS equations were interpreted as giant gravitons \cite{Bena:2005}. In \cite{BFM:2009} we have shown 
that the coupled BPS equations collapse effectively into a single-family BPS equation at special loci in parameter space of the two family model. A similar observation was also made concerning the more general coupled non-BPS variational equations. All this was the consequence of the invariance of the 
collective-field Hamiltonian of the two-family Calogero model under an Abelian group of strong-weak-coupling dualities \cite{BFM:2007}, analogous to (\ref{duality}). 
In \cite{BMPRD:2007} and \cite{BMMPRD:2009} it was shown that a large class of solitons in the two-family Calogero model can be 
obtained by reducing it to the effectively one-family Calogero model.

These results on the collective-field solitons in various variants of the two-family Calogero
model motivated us to revisit the collective-field theory of the original single-family Calogero model in \cite{BFM:2009}, where we studied, among
other things, the periodic soliton crystal solutions originally discovered in \cite{Polychronakos:1994xg}.  In this paper we focus on a particular type of such soliton crystals - 
the periodic BPS density waves of the collective field hamiltonian, and study their quantum stability. We present complete analysis of quadratic fluctuations around these BPS solutions.  The corresponding fluctuation hamiltonian (i.e., the stability operator) is diagonalized in terms of bosonic creation and annihilation operators which correspond to 
the complete orthogonal set of Bloch-Floquet eigenstates of a related periodic Schr\"odinger hamiltonian, which we derive explicitly. The resulting fluctuation spectrum is 
positive, and therefore these density waves are stable.

This paper is organized as follows: In Section 2 we review the solution of  the static BPS equation associated with the collective Hamiltonian (\ref{Hcollective}). 
This is done by converting it into a Riccati equation which can then solve explicitly. The solution is a static
periodic soliton - the finite amplitude wave solution of
\cite{Polychronakos:1994xg}.  

In Section 3 we study quadratic fluctuations around these BPS density waves and discuss their quantum-mechanical stability as was described in the paragraph next to 
previous.  The complete orthogonal set of Floquet-Bloch eigenstates of the related periodic Schr\"odinger hamiltonian, in terms of which we diagonalize the fluctuation hamiltonian, are derived explicitly in Section 4. Remarkably, the fluctuation spectrum is independent of the parameter which determines the density wave's amplitude. As a consequence, the sum over zero-point 
energies of the field-theoretic fluctuation hamiltonian, and its ensuing normal-ordering and regularization, are the same as in the case of fluctuations around constant 
density background, namely, the ground state. We close by making some brief remarks concerning fluctuations around non-BPS density waves in Section 5. 
Finally, some technical details of the collective field formalism are relegated to the Appendix.

\section{Review of periodic BPS solutions of the collective field equation }
\label{section2}
\setcounter{footnote}{1}

The Hamiltonian (\ref{Hcollective}) is essentially the sum of two
positive terms. Its zero-energy classical solutions are zero-momentum, and therefore
time independent configurations of the collective field
(\ref{collective}), which are also solutions of the BPS equation
\beq\label{BPS} B[\rho] \equiv \frac{\lambda - 1}{2}
\frac{\partial_{x} \rho}{\rho} + \lambda \pv \int \frac{ dy
\rho(y)}{x - y} = 0\,. \eeq

It is easy to check that the duality transformation
(\ref{duality}) maps a solution $\rho(x)$ of (\ref{BPS}) with
coupling $\lambda$ onto another solution $\tilde\rho(x) =
-\lambda\rho(x)$ of that equation with coupling $\tilde\lambda =
{1\over\lambda}$. As we shall see below in Eq. (\ref{rhorhoH}),
all solutions of (\ref{BPS}) are of definite sign, and never
vanish along the real axis. Thus, such a positive solution of
(\ref{BPS}) is mapped by (\ref{duality}) onto a negative solution,
and vice-versa.

The BPS equation (\ref{BPS}) may be written alternatively as
\beq\label{BPS1} (\lambda -1)\,\partial_x\rho =
2\pi\lambda\rho\rho^H\,,\eeq where $\rho^H$ is the
Hilbert-transform 
\beq\label{Hilbert} \rho^H(x) = {1\over \pi}
\pv\int_{-\infty}^\infty \,dy\, {\rho (y) \over y - x} \eeq
of $\rho$. Note that for
$\lambda = 1$, where the CM describes non-interacting fermions,
the {\em only} solution of (\ref{BPS1}) is $\rho = \rho_0 = {\rm
const.}$ This is also the case at the bosonic point $\lambda=0$. Henceforth, we shall assume $\lambda\neq 0,1$.
Space independent constant configurations  $\rho = \rho_0$ are obviously solutions of (\ref{BPS}) also for  $\lambda\neq 0,1$. 
However, for such values of $\lambda$, (\ref{BPS}) bears also the periodic space-dependent density wave solutions, whose derivation we review in this section.  

The proper way to solve this nonlinear integro-differential
equation is to consider it together with its
Hilbert-transform\cite{abanov, AJJ:2006, BFM:2009} \beq\label{BPS-H} (\lambda
-1)\,\partial_x\rho^H = \pi\lambda ((\rho^H)^2 - \rho^2 +
\rho_0^2)\,,\eeq where on the RHS we used the identity\footnote{\label{HT} For a 
compendium of useful identities involving Hilbert-transforms see Appendix A of \cite{BFM:2009} and also Appendix A of the second paper cited in \cite{abanov}.} 
\beq\label{ffHilbert}
2(\rho\rho^H)^H = (\rho^H)^2 - \rho^2 + \rho_0^2\, \eeq
(and the fact that $\partial_x\rho^H =
(\partial_x\rho)^H$ on the LHS). Here $\rho_0$ is a real parameter
such that \beq\label{subtraction} \int\limits_{-\infty}^\infty\,
dx\, (\rho(x) - \rho_0) = 0\,.\eeq It arises from the fact that we
seek a solution of $\rho (x)$ which need not necessarily decay at
spatial infinity. Note that (\ref{BPS-H}) is even in $\rho_0$. By
definition, the sign of $\rho_0$ coincides with that of $\rho
(x)$, the solution of (\ref{BPS-H}). A positive solution $\rho
(x)\geq 0$ corresponds to a BPS configuration of particles, and a
negative one, to a configuration of antiparticles, as was
mentioned following (\ref{duality}).

We proceed as follows. Given the density $\rho(x)$, consider the
resolvent \beq\label{Phi} \Phi(z) =
{1\over\pi}\int\limits_{-\infty}^\infty \,dy\,{ \rho(y)\over y-z}
\eeq associated with it, in which $z$ is a complex variable.

The resolvent $\Phi(z)$ is evidently analytic in the complex
plane, save for a cut along the support of $\rho (x)$ on the real
axis. From the identity 
\beq\label{PS} {1\over x\mp i0} = {P\over x} \pm
i\pi\delta (x)\,, \eeq 
we obtain \beq\label{Phipm}
\Phi_\pm (x) \equiv \Phi(x\pm i0) = \rho^H(x) \pm i\rho(x) \,.
\eeq Thus, if $\Phi (z)$ is known, $\rho (x)$ can be determined from the discontinuity
of $\Phi (z)$ across the real axis.

An important property of $\Phi(z)$, which follows directly from
the definition (\ref{Phi}), is \beq\label{herglotz0} \Im\, \Phi(z)
= {\Im\,z\over\pi}\,\int\limits_{-\infty}^\infty \,
{\rho(y)\,dy\over |z-y|^2}\,. \eeq Thus, if $\rho(x)$ does not
flip its sign throughout its support, we have \beq\label{herglotz}
\sign\,\left(\Im\, \Phi(z)\right) =
\sign\,\left(\Im\,z)\right)\sign\,\left(\rho (x)\right)\,.\eeq We
shall use this property to impose certain further conditions on
the solution of (\ref{Riccati1}) below.

It follows from (\ref{Phipm}) that (\ref{BPS1}) and (\ref{BPS-H})
are, respectively, the imaginary and real parts of the  Riccati
equation \beq\label{Riccati} (\lambda -1)
\partial_x \Phi_\pm (x) = \pi\lambda (\Phi_\pm^2(x) + \rho_0^2)
\eeq obeyed by both complex functions $\Phi_\pm (x)\,.$ Let
$\Phi_\pm (z)$ be the analytic continuations of $\Phi_\pm (x)$
into the $z-$upper and lower half planes, respectively. These
functions are evidently the two solutions of \beq\label{Riccati1}
(\lambda -1)\partial_z \Phi (z) = \pi\lambda (\Phi (z)^2 +
\rho_0^2)\,, \eeq subjected to the boundary conditions $\Phi^*_+
(x+i0) = \Phi_- (x-i0)$ and $\sign\left(\Im \Phi_+ (x + i0)\right)
= \sign\left(\rho(x)\right) = \sign\rho_0,$ from (\ref{Phipm}).
The resolvent (\ref{Phi}) is then obtained by patching together
$\Phi_+(z)$ in the upper half-pane and $\Phi_-(z)$ in the lower
half-plane.

The standard way to solve (\ref{Riccati1}) is to write it as
\beq\label{Riccati2} \left({1\over \Phi(z) - i\rho_0} - {1\over
\Phi(z) + i\rho_0} \right)\,\partial_z \Phi (z) = i k \,, \eeq
where \beq\label{k} k = {2\pi\lambda\rho_0\over\lambda -1}\,, \eeq
is a real parameter.

Straightforward integration of (\ref{Riccati2}) then yields the
solutions \beq\label{PhiSol} \Phi_\pm (z)  = i\rho_0\,{1 + e^{ikz
-u_\pm}\over 1 - e^{ikz -u_\pm}}\,,\eeq where $u_\pm$ are
integration constants. The boundary condition $\Phi_+^* (x+i0) =
\Phi_- (x-i0)$ then tells us that $u_- = -u_+^*$. Clearly, $\Im
u_+$ can be absorbed by a shift in $x$. Therefore, with no loss of
generality we set $\Im u_+ = 0$. The second boundary condition
$\sign\left(\Im \Phi_+ (x+i0)\right) = \sign\rho_0 $ then tells us
that $u \equiv \Re u_+ >0\,.$ Thus, $\Phi_\pm (z)$ are completely
determined and we obtain (\ref{Phi}) as \beq\label{PhiSolFinal}
\Phi (z)  = i\rho_0\,{1 + e^{ikz -u\,\sign(\Im z)}\over 1 - e^{ikz
-u\,\sign(\Im z)}}\,.\eeq As can be seen in (\ref{rhorhoH}) below,
the density $\rho(x)$ associated with (\ref{PhiSolFinal}) is
indeed of definite sign, namely, $\sign\rho_0$.

The asymptotic behavior of (\ref{PhiSolFinal}) is such that
\beq\label{asymptotics} \Phi (\pm i\infty ) = \pm i\rho_0\,\sign
k\,. \eeq This must be consistent with (\ref{herglotz}), which
implies (together with the fact that $\sign\left(\rho(x)\right) =
\sign\rho_0$) that $k$ must be {\em positive}. In other words, as
can be seen from (\ref{k}), positive BPS density configurations
($\rho_0 > 0 $) exist only for $\lambda
>1$, and negative BPS densities ($\rho_0 < 0 $) arise only for
$0<\lambda<1$. The duality symmetry (\ref{duality}), which
interchanges the domains $0< \lambda <1$ and $\lambda >1$, maps
these two types of BPS configurations onto each other.
Positivity of $k$, the condition (\ref{positivity}) and the fact that $\sign\rho(x)=\sign\rho_0$ imply 
that we always have 
\beq\label{signs}
\sign \rho_0 = \sign m = \sign (\lambda -1)\,.
\eeq

Now that we have determined $\Phi (z)$, let us extract from it the
BPS density $\rho (x)$ and its Hilbert transform $\rho^H(x)$. From
(\ref{PhiSolFinal}) we find that \beq\label{PhiSolplus} \Phi_+(x)
= \Phi(x+i0) = \rho_0\, {-\sin\, kx + i\sinh u\over \cosh u -\cos
kx}\,,\eeq from which we immediately read-off the solution of the
BPS-equation (\ref{BPS}) as \beqra\label{rhorhoH} \rho (x) &=&
\,\,\,\,\rho_0\, {\sinh u\over \cosh u -\cos
kx}\nonumber\\{}\nonumber\\\rho^H(x) &=&  - \rho_0\, {\sin kx
\over \cosh u -\cos kx}\,,\eeqra where both $k > 0$ and $u>0$, and
the sign of $\rho(x)$ coincides with that of $\rho_0$. That
$\rho^H$ in (\ref{rhorhoH}) is indeed the Hilbert-transform of
$\rho$ can be verified by explicit calculation.

The static BPS-soliton, given by $\rho(x)$ in (\ref{rhorhoH}), is
nothing but the finite-amplitude solution of
\cite{Polychronakos:1994xg}. It comprises a two-parameter family
of spatially periodic solutions, all of which have zero energy
density, by construction. The period is \beq\label{period} T =
{2\pi\over k} = {\lambda -1\over\lambda\rho_0}\,.\eeq It can be
checked by explicit calculation\footnote{The best way to do this
computation is to change variables to $t = e^{ikx}$ and transform
the integral into a contour integral around the unit circle.} that
\beq\label{period-av} {1\over T}\int\limits_{\rm period}\, \rho
(x)\, dx = \rho_0\,,\eeq and therefore that
$\int\limits_{-\infty}^\infty\, (\rho (x) - \rho_0) \, dx = 0\,,$
as required by definition of $\rho_0$. Thus, the parameter
$\rho_0$ determines both the period of the solution $\rho (x)$, as
well as its period-average, and the other (positive) parameter $u$
determines the amplitude of oscillations between its extremal values
\beq\label{extrema}
\rho_{min} = \rho_0\tanh {u\over 2}\quad  {\rm and}\quad   \rho_{max} = \rho_0\coth {u\over 2}\,.
\eeq
Note also from (\ref{period-av}), that the number of particles per
period is \beq\label{ppp} T\rho_0 = {\lambda-1\over
\lambda}\,.\eeq

A few limiting cases of (\ref{rhorhoH}) are worth
mentioning. Thus, if we let $u\rightarrow 0\,,$ we obtain a comb
of Dirac $\delta-$functions \beq\label{comb} \rho (x) =
{\lambda-1\over \lambda}\, \sum\limits_{n\in Z\!\!\!Z}\, \delta (x
- nT)\,.\eeq If, in addition to $u\rightarrow 0$, we also let $k$
tend to zero (or equivalently, let the period $T$ diverge), such
that $b = {u\over k}$ remains finite, we obtain the BPS soliton
solution \cite{Polychronakos:1994xg, Andric:1994nc}
\beq\label{lump} \rho (x) = {\lambda-1\over
\lambda}\,{1\over\pi}\, {b\over b^2 + x^2}\,.\eeq In fact, the
original construction of the periodic soliton (\ref{rhorhoH}) in
\cite{Polychronakos:1994xg} was done by juxtaposing infinite
solitons (\ref{lump}) in a periodic array.

Note that the relation (\ref{ppp}) is preserved in both limiting
cases discussed above, since the RHS of (\ref{ppp}) depends neither on $u$ nor on
$k$.

Finally, by letting $u\rightarrow\infty$ in (\ref{rhorhoH}), we obtain the uniform solution $\rho=\rho_0$ of (\ref{BPS}), independently of $k$. 
As can be seen from (\ref{k}),  $k$ blows up as $\lambda\rightarrow 1$, namely, at the point of non-interacting fermions. Consequently, in this limit, the BPS density wave
oscillates wildly between its extrema (\ref{extrema}). 
This is clearly a pathological situation, unless these extrema coincide, which happens only when $u\rightarrow\infty$, namely, the uniform configuration 
$\rho=\rho_0$. The latter is the only solution of (\ref{BPS}) at $\lambda=1$. Thus, if we seek match smoothly our rapidly oscillating solutions in the {\em vicinity} of 
$\lambda = 1$ with the constant solution $\rho=\rho_0$  precisely at $\lambda=1$, we have to insist that the amplitude of the density wave should vanish as 
$\lambda\rightarrow 1$ , i.e., let $u\rightarrow\infty$. More quantitatively, according to (\ref{extrema}), a typical estimate
of a derivative of the density wave (\ref{rhorhoH}) is $k(\rho_{max}-\rho_{min}) = {2k\rho_0\over\sinh u}$, which for large $k$ and $u$ behaves asymptotically 
like $ke^{-u}\sim {e^{-u}\over \lambda-1}$. Thus, we must have $e^u\gg {1\over\lambda-1}$  in (\ref{rhorhoH}), in order for the density waves 
to cross-over smoothly to the uniform solution at $\lambda=1$. None of these complications arise at $\lambda=0$, corresponding to non-interacting bosons. In this limit 
$k\rightarrow 0$ independently of $u$, and (\ref{rhorhoH}) tends to a uniform solution trivially.

\section{Fluctuations}
\label{section3}
\setcounter{footnote}{1}

The collective-field formalism provides a systematic framework for
the $ {1\over N}$ expansion \cite{Andric:1988vt, ABJ:1995, Andric:1991wp}.
By expanding the collective Hamiltonian (\ref{Hcollective}) around
the static BPS solution in (\ref{rhorhoH}), which we shall henceforth denote by $\rho_s(x)$,  we can go beyond the leading order and
obtain the spectrum of low-lying excitations above $\rho_s(x)$.
Here we shall concentrate on the next-to-leading terms, embodied
in the quadratic fluctuations around $\rho_s(x)$.

To this end we write \beq\label{fluct} \rho(x) = \rho_s(x) + \eta (x)\,,\eeq
where $\eta (x)$ is a small density fluctuation around the wave
solution $\rho_s(x)$, being typically of order $1/N$.  Similarly, we shift $\mu=\mu_s + \delta\mu$, where 
$\mu_s = 0$ is the value of the chemical potential corresponding to $\rho_s(x)$.
Due to (\ref{conservation}), which is already satisfied by $\rho_s(x)$, we clearly  must have $\int\,dx \,\eta = 0\,.$ This latter constraint 
is enforced by the shifted chemical potential $\delta\mu$, as can be seen in (\ref{Hcollbps1}) below.  

A convenient intermediate step is to expand the BPS combination $B[\rho]$ around $\rho_s(x)$. We obtain 
\beq\label{Bfluct} B[\rho_s + \eta] = \left[ {\lambda-1\over 2}\partial_x\left(\eta\over \rho_s\right) - \lambda\pi\eta^H\right] 
- {\lambda-1\over 2}\partial_x\left(\eta\over \rho_s\right)^2 +\ldots\,,\eeq
where the ellipsis stands for terms cubic in $\eta$ and higher, and where we used the BPS equation $B[\rho_s(x)]=0$, (\ref{BPS}).
Then, substituting (\ref{Bfluct}) in (\ref{Hcollective}), and expanding $H_{coll}$ 
to second order in $\eta(x)\,,\pi (x)$ and $\delta\mu$,  we obtain the quadratic fluctuation hamiltonian $H_{coll}^{(2), BPS} $ around $\rho_s(x)$ as 
\beqra\label{Hcollbps1}
H_{coll}^{(2),BPS}& =&  {1\over 2m}\,\int\,dx\, \rho_s(x)\,\left(\partial_x\pi(x)\right)^2 + 
{1\over 2m}\,\int\,dx\, \rho_s(x)\,\left[{\lambda-1\over 2}\partial_x\left(\eta\over \rho_s\right) - \lambda\pi\eta^H\right]^2\nonumber\\{}\nonumber\\
& -&\delta\mu\,\int\,dx \,\eta + H_{sing} \,.\eeqra
Here we used the constraint (\ref{conservation}) and the BPS equation (\ref{BPS}) to eliminate all terms linear in the shifted 
quantities $\eta$ and $\delta\mu$.  

Let us now evaluate $H_{sing}$ in the background of $\rho_s(x)$. This will be also useful for later reference. 
The singular term multiplying $\rho(x)$ in the first term in (\ref{Hsing}) is a constant, $C$. We can evaluate this constant as a divergent integral in 
momentum space. To this end, note that according to (\ref{absp-me}) below, $C$ is proportional to the singular matrix element $\langle x| |p| |x\rangle$. Thus\footnote{This integral should, of course, be cut-off at some large momentum. The point is that the same regularization should be used for both the uniform and density wave backgrounds.}
\beq\label{C-divergent}
C = -{\lambda\over 2m}\left.\partial_x{P\over x-y} \right |_{y=x}  = -{\lambda\pi\over\ 2m}\langle x| |p| |x\rangle = -{(\lambda-1)k\over 4m\rho_0}\int\,{dq\over2\pi}\,|q|\,,
\eeq
where we have also used (\ref{k}). Since $C$ is constant, according to (\ref{subtraction}) we may replace $\rho_s(x)$ under the integral in (\ref{Hsing}) by $\rho_0$. 
Therefore,  $H_{sing}$ in this background is {\em the same as in the familiar constant background} $\rho(x)=\rho_0$.   
Combining $\int\,dx\,C\rho_0$, the first term in (\ref{Hsing}), with the momentum space representation of the second term there, 
we obtain our desired result as 
\beq\label{Hsing-Fourier}
H_{sing}  = \left(\int dx\right)\,\int\,{dq\over 2\pi}\,{\lambda-1\over 4m}\,(q^2 - k|q|)\,,
\eeq
We shall make use of this explicit form later. 

The hamiltonian (\ref{Hcollbps1}) is essentially the sum of squares of local hermitian operators, each of which is multiplied by the positive function ${\rho_s(x)\over m} $. 
Thus, the expectation value of (\ref{Hcollbps1}) with respect to any (normalizable) wave functional $\Psi[\eta]$ is positive, for all values of $\lambda $. 
The fluctuation spectrum about $\rho_s(x)$ is therefore {\em positive}, and the BPS density waves $\rho_s(x)$ are {\em stable}. This, of course, comes at no surprise, 
since the energy density tied with $\rho_s(x)$ is strictly zero - the lowest possible value for (\ref{Hcollective}).

This positivity of (\ref{Hcollbps1}), i.e., its  quadratic structure, clearly calls for the introduction of the operator 
\beq\label{Aop}
A(x) = \pa_x \pix+ i\left[{\lambda-1\over 2}\partial_x\left(\eta\over \rho_s\right) - \lambda\pi\eta^H\right]
\eeq
along with its hermitian adjoint 
\beq\label{Aopdagger}
A^\dgg(x) = \pa_x\pix - i\left[{\lambda-1\over 2}\partial_x\left(\eta\over \rho_s\right) - \lambda\pi\eta^H\right]\,.
\eeq
It follows from the canonical commutation relations 
\beq\label{CCR}
[\eta(x),\pi(y)] = i\delta(x-y)\,,\quad  [\eta(x),\eta(y)] =[\pi(x),\pi(y)] = 0
\eeq
that $A$ and $A^\dagger$  satisfy the commutation relation  
\beq\label{Acommut}
\left[A(x)\,,\,A^\dgg(y)\right] = (1-\lambda)\,\pa_x\pa_y\,\left({\delta(x-y)\over \rho_s(x)}\right)  + 2\lambda\pa_x\,{P\over x-y}\,,
\eeq
with all other commutators vanishing.

In terms of the operators (\ref{Aop}) and (\ref{Aopdagger}), we may write the quadratic hamiltonian (\ref{Hcollbps1}) as 
\beqra\label{Hcollbps2}
H_{coll}^{(2),BPS}& =&  {1\over 2m}\,\int\,dx\, \rho_s(x)\,A^\dgg(x)\,A(x)   \nonumber\\{}\nonumber\\ 
&+&  {1\over 4m}\,\int\,dx\, \rho_s(x)\,\left[A(x)\,,\,A^\dgg(x)\right] + H_{sing} 
 -\delta\mu\,\int\,dx \,\eta  \,.\eeqra
As can be clearly seen by comparing (\ref {Acommut}) and (\ref{Hsing}), the commutator term in (\ref {Hcollbps2}) would cancel the singular term $H_{sing}$ if $\rho_s$ were constant, in accordance with (\ref{Hsing-Fourier}). We shall see below (see (\ref{commutator-term})) that this commutator term, in fact, exactly cancels the expression in (\ref{Hsing-Fourier}) also in the background $\rho_s(x)$.  Thus, we obtain\footnote{The term $\delta\mu\,\int\,dx \,\eta $ merely constrains the zero-momentum Fourier mode of $\eta(x)$ to vanish, and we shall henceforth not write it explicitly} 
\beq\label{Hcollbps}
H_{coll}^{(2),BPS} =  {1\over 2m}\,\int\,dx\, \rho_s(x)\,A^\dgg(x)\,A(x)\,.\eeq
As we shall see below, this hamiltonian requires one last zero-point energy subtraction to render it finite.

\subsection{Diagonalization of $H_{coll}^{(2),BPS}$ in terms of bosonic creation and annihilation operators } 
\label{section3-1}
Our goal is to diagonalize the manifestly positive hamiltonian (\ref{Hcollbps}). This could be achieved by decomposing $\sqrt{\rho_s(x)\over 2m} A(x)$ and $\sqrt{\rho_s(x)\over 2m} A^\dagger(x)$ into orthonormal modes. To this end, let us first observe that we may rewrite (\ref{Acommut})  as
\beq\label{Acommut-h}
\sqrt{\rho_s(x)\over 2m}\left[A(x)\,,\,A^\dgg(y)\right] \sqrt{\rho_s(y)\over 2m}= {\bf 1}_{Fock}\cdot\langle x| H| y\rangle\,,
\eeq
where $H$ is the single-particle hamiltonian
\beq\label{H}
H =  (1-\lambda)\,\sqrt{\rho_s(x)\over 2m}\left[  p {1\over\rho_s(x)}p - {k\over \rho_0} |p|\right]\sqrt{\rho_s(x)\over 2m}\,,
\eeq
in which $x$ and $p$ are canonically conjugate position and momentum operators. The first term in (\ref{H}), quadratic in $p$, can be read-off from (\ref{Acommut}) 
in a straightforward manner. In fact, it has the standard non-relativistic hamiltonian form, since it is proportional to
\beqra\label{standard}
\left({1\over\sqrt{\rho_s}}p\sqrt{\rho_s}\right)^\dagger \left({1\over\sqrt{\rho_s}}p\sqrt{\rho_s}\right)&=& p^2 + \left({1\over 2}\partial_x\log\rho_s(x)\right)^2 -
{1\over 2}\partial_x^2\log\rho_s(x)\nonumber\\{} &=& p^2+{k^2\over 4}\left[\left({\rho_s(x)\over\rho_0}\right)^2-1\right]\,,
\eeqra
where in the last step we used the BPS equation (\ref{BPS}). The second term in (\ref{H}) arises because the matrix-element
\beq\label{absp-me}
\langle x | |p| |y\rangle  = {1\over \pi} \partial_x {P\over x-y}\,,
\eeq
which follows from the identity\footnote{This identity can be easily established by applying the operator $|p|$ to the Fourier integral representation of  $\psi (x)$, and using  the formula $(e^{ipx})^H = i e^{ipx}\sign p$.} 
\beq\label{absp}
|p|\psi (x) = -\partial_x(\psi^H(x))\,.
\eeq 
In addition, in deriving the second term in (\ref{H}), one has to invoke equations (\ref{BPS}) and (\ref{k}).

The operator $H$ is therefore a hermitian periodic Schr\"odinger operator, with period $T = {2\pi\over k}$ (Eq.(\ref{period})). Consequently, it has a complete set of orthogonal 
Floquet-Bloch eigenstates  $\vphi_{q}(x)$ and corresponding energy eigenvalues $\omega(q)$, where $q$ is quasi-momentum. 
Note that $H$ is {\em not} positive definite. Hence $\omega(q)$ may become negative over some range of $q$.

As we shall show in Section 4 below, {\em the spectrum of $H$ has no gaps!} This peculiar feature is obviously the result of the unconventional $|p|$ term in $H$. 
Since there are no gaps in the spectrum, we shall take the quasi-momentum $q$ to range from $-\infty$ to $+\infty$ (i.e., work in the extended-Brillouine zone scheme).

We shall now prove that  the complete orthogonal set of Floquet-Bloch eigenstates $\vphi_{q}(x)$ can be used to diagonalize the fluctuation hamiltonian
(\ref{Hcollbps}) over an appropriate Fock space. This can be done exclusively on basis of the general properties of the periodic Schr\"odinger hamiltonian $H$, without any
reference to the explicit form of the eigenstates $\vphi_{q}(x)$. We shall therefore defer derivation of the explicit form of the eigenstates $\vphi_{q}(x)$ and their corresponding energy eigenvalues $\omega(q)$ to Section 4. 

Let us now list a few standard facts about the spectrum of $H$. By definition, the Floquet-Bloch eigenstates are quasi-periodic, and satisfy
\beq\label{floquet}
\vphi_{ q}(x + T) = e^{iqT}\vphi_{ q}(x)\,.
\eeq
The completeness relation $\int\limits_{-\infty}^\infty\,dq \,|\omega(q)\rangle\langle \omega(q)|\, = 1\!\!1,$ obeyed by the Floquet-Bloch eigenstates, may be written as 
\beq\label{completeness2}
\int\limits_{-\infty}^\infty\,dq \,\vphi_{ q}(x)\,\vphi_{ q}^*(y)  = \delta (x-y)\,.
\eeq
Here, as usual, $\langle x|\omega(q)\rangle  = \vphi_{ q}(x)$\,.
Orthogonality is expressed as 
\beq\label{orthogonality}
\langle \omega(q)|\omega (q')\rangle = \int\limits_{-\infty}^\infty\, dx\, \vphi_{ q}^*(x) \vphi_{ q'}(x) = \delta(q-q')\,.
\eeq
Finally, using the spectral decomposition $H = \int\limits_{-\infty}^\infty\, dq\,  \omega(q)\,|\omega(q)\rangle\langle \omega(q)|$ we obtain the matrix element of $H$ in the position basis as 
\beq\label{decomposition1}
\langle x|H|y\rangle  = \int\limits_{-\infty}^\infty\, dq\,  \omega(q)\,\vphi_{ q}(x) \,\vphi_{q}^*(y) \,.
\eeq

Let us revert back to the discussion leading from (\ref{Hcollbps2}) to (\ref{Hcollbps}), namely cancellation of (\ref{Hsing-Fourier}) by the commutator term in (\ref{Hcollbps2}). 
By combining (\ref{Acommut-h}), (\ref{orthogonality})  and (\ref{decomposition1}) we can express that commutator term as 
\beq\label{commutator-term}
{1\over 4m}\,\int\,dx\, \rho_s(x)\,\left[A(x)\,,\,A^\dgg(x)\right]  = {1\over 2}\,\int\,dx\, \langle x|H|x\rangle  = \left(\int dx \right)\, \int\limits_{-\infty}^\infty\, {dq\over 2\pi}\,  {\omega(q)\over 2} \,.
\eeq
In other words, this commutator term is nothing but the zero-point energy of  a non-interacting bosonic field theory with eigenmodes given by the eigenvalues of $H$ in 
(\ref{H}), $\om(q)$. We compute this dispersion relation in the next section. It is given by (\ref{omega-final}). Substituting (\ref{omega-final}) in (\ref{commutator-term}) we see that 
it exactly cancels $H_{sing}$ in (\ref{Hsing-Fourier}), as promised. 
   
Since we seek a diagonalization of the fluctuation hamiltonian (\ref{Hcollbps}) over an appropriate Fock space, we shall associate with each Floquet-Bloch eigenstate
$\vphi_{q}(x) $ a pair of bosonic creation and annihilation operators
$a(q), a^\dagger (q)$ which obey the standard bosonic commutator algebra 
\beqra\label{bosonic}
\left [ \, a(q)\, ,\, a^\dgg(q')\,\right] &=& \delta (q-q')\,,\nonumber\\{}\nonumber\\
\left [ \, a(q)\,,\, a(q') \,\right ] &=& \left [ \, a^\dgg(q)\,,\, a^\dgg(q') \,\right ]  = 0\,.
\eeqra
It is then a matter of straightforward calculation to show that the algebra (\ref{Acommut}) is realized by the normal-mode expansions 
\beqra\label{AAdagger}
\sqrt{\rho_s(x)\over 2m}\,A(x)   &=& \int\limits_{-\infty}^\infty\,dq\, |\omega(q)|^{1\over 2} \vphi_{q}(x)\,\left[\theta\left(\omega(q)\right)a(q) + \theta\left(-\omega(q)\right)a^\dagger(q) \right]\,,\nonumber\\{}\nonumber\\
\sqrt{\rho_s(x)\over 2m}\,A^\dagger(x)   &=& \int\limits_{-\infty}^\infty\,dq\, |\omega(q)|^{1\over 2} \vphi^*_{q}(x)\,\left[\theta\left(\omega(q)\right)a^\dagger(q) + \theta\left(-\omega(q)\right)a(q) \right]\,,\nonumber\\{}
\eeqra
which take into account the fact that $\om(q)$ flips its sign. Indeed, it is easy to see that (\ref{AAdagger}) are consistent with $[A(x),A(y)] =0$. In addition, fulfillment of (\ref{Acommut-h}), namely, 
\beq\label{AAdaggercom}
\left[\sqrt{\rho_s(x)\over 2m} A(x)\,,\sqrt{\rho_s(y)\over 2m} A^\dagger (y)\right]  = 
\langle x| H|y\rangle\,,
\eeq
follows from (\ref{AAdagger}), (\ref{bosonic}), (\ref{decomposition1}), (\ref{orthogonality}) and (\ref{H}). 
Thus, (\ref{AAdagger}) is a legitimate realization of the operators $A(x)$, $A^\dagger(x)$. 
Finally, upon substituting the normal mode expansions (\ref{AAdagger}) in (\ref{Hcollbps}) we obtain, using the orthogonality relation (\ref{orthogonality}), the desired {\em diagonal} form of $H_{coll}^{(2),BPS} $ as\footnote{Here we used $\theta^2(\om) = \theta (\om)$, as well as $\theta(\om)\theta(-\om) = 0$.  The latter is responsible for the absence of $aa$ and $a^\dagger a^\dagger$ terms on the RHS of (\ref{Hcollbps1n}).} 
\beq\label{Hcollbps1n}
H_{coll}^{(2),BPS} =  \int\limits_{-\infty}^\infty\,dq\,|{\omega(q)}| \left[ \theta\left(\omega(q)\right)a^\dagger(q) a(q) + 
\theta\left(-\omega(q)\right) a(q)a^\dagger(q) \right]\,.
\eeq
This manifestly positive operator is not normal-ordered, due to the expansions (\ref{AAdagger}), which contain both creation and annihilation operators. 
We should perhaps mention that a similar situation arises in the two dimensional anyonic model studied in \cite{Andric:1993xm}. The divergent expectation value 
of (\ref{Hcollbps1n}) in the Fock vacuum is given by the sum over zero-point energies of negative $\omega(q)$ modes, namely, 
\beqra\label{zpe-BPS}
\langle 0 | H_{coll}^{(2),BPS} | 0\rangle &=& \left(\int dx \right)\, \int\limits_{-\infty}^\infty\,{dq\over 2\pi}\,|{\omega(q)}|  \theta\left(-\omega(q)\right)  = - \left(\int dx \right)\,\int
\limits_{|q|>k}\,{dq\over 2\pi}\,{\omega(q)}\nonumber\\{}\nonumber\\ &=& 2H_{sing} + \left(\int dx \right)\,{\lambda-1\over 12\pi m}k^3\,,
\eeqra
where in the last step we used (\ref{Hsing-Fourier}) and (\ref{omega-final}).

Note that the Floquet-Bloch eigenvalue $\om(q)$ given in (\ref{omega-final}) coincides with the dispersion relation of quasi-particles in the presence of uniform 
condensate $\rho_0$, namely, the ground state. Thus, according to (\ref{zpe-BPS}), quantum corrections to the energy density tied with the BPS-density wave studied here, due to  quadratic fluctuations of the collective field, 
coincide with the analogous corrections around the uniform condensate background. In other words, as one starts from the uniform solution of (\ref{BPS}), namely, 
(\ref{rhorhoH}) taken at infinite $u$, and then reduces $u$ continuously to some positive finite value, there are no (next-to-leading order) quantum corrections to the energy density tied with the BPS-density wave relative to the ground state. (The classical values of energies in both cases, are of course null.) 
Moreover, since the single BPS soliton, or lump, is obtained as the limiting case (\ref{lump}) of the BPS-density wave, its mass it not corrected by quadratic fluctuations either. 

Finally, using the commutator algebra (\ref{bosonic}) to normal-order (\ref{Hcollbps1n}), and subtracting the divergent contribution (\ref{zpe-BPS}), we obtain the desired 
diagonalized fluctuation hamiltonian simply as 
\beq\label{Hcollbpsfinal}
{\bf :}\left(H_{coll}^{(2),BPS}\right) {\bf :} \,  =  \int\limits_{-\infty}^\infty\,dq\,|{\omega(q)}|  \,a^\dagger(q)a (q)\,.
\eeq
It is manifestly positive definite, as required. We need not worry about positivity of $\omega(q)$ since only its absolute value enters (\ref{Hcollbpsfinal}).

\section{The exact Floquet-Bloch eigenfunctions and energy band}
\label{section4}
\setcounter{footnote}{1}

Now that we have diagonalized $H_{coll}^{(2),BPS}$, it remains to determine the Floquet-Bloch eigenstates $\vphi_{q}(x)$ and corresponding eigenvalues $\omega(q)$ 
explicitly, and establish their orthogonality and completeness.  

The eigenvalue equation $H\phi_{q} (x)= \omega(q)\phi_{q}(x)$ may be written explicitly as 
\begin{equation}\label{ev1}
(1-\lambda)\sqrt{\rho_s\over 2m} \,p {1\over \rho_s} p\, (\sqrt{\rho_s\over 2m}\phi) + 2\pi\lambda\sqrt{\rho_s\over 2m}\,|p|\,(\sqrt{\rho_s\over 2m}\phi) = \omega\phi\,,
\end{equation}
where we have suppressed any $q$-dependence for brevity. Note also the slight change in notation of eigenstates from $\vphi_q$ to $\phi_q$. (This is done in order to avoid possible 
confusion in the discussion below, and we shall return to the original notation toward the end of this section.) This equation clearly calls for defining a new unknown function 
\begin{equation}\label{psidef}
\psi = \sqrt{\rho_s\over 2m}\,\phi
\end{equation}
which has the same quasi periodicity (\ref{floquet}) as $\phi$. In terms of $\psi$ we have
\begin{equation}\label{ev2}
\left({1-\lambda\over 2m}\right)\rho_s \,p {1\over \rho_s} p\,\psi + 2\pi\lambda\left({\rho_s\over 2m}\right)|p|\,\psi = \omega\psi\,.
\end{equation}
Recalling (\ref{absp})  and the BPS equation (\ref{BPS}), 
we may write (\ref{ev2}) more explicitly as
\begin{equation}\label{psieq}
(\lambda -1)\partial_x^2 \psi -2\pi\lambda\left[\rho_s^H\partial_x\psi + \rho_s \partial_x\psi^H\right] = 2m\omega\psi\,.
\end{equation}

\subsection{Solving Schr\"odinger's Equation}

Equation  (\ref{psieq}) is very suggestive of taking its Hilbert transform, using the identity\footnote{For derivation of (\ref{Hidentity}) for real functions see Appendix A of \cite{BFM:2009}. For a derivation of (\ref{Hidentity}) for complex valued functions analytic in one half-plane, where they also decay at infinity (e.g., exponentials $e^{ipx}$), see Appendix A in the second paper in \cite{abanov}.   If $f$ and $g$ are of the latter type, they are eigenfunctions of the Hilbert transform, with eigenvalue $+i$ for analyticity in the UHP, and $-i$ for the LHP. For such functions $f_0=g_0=0$ due to Cauchy's theorem.} 
\begin{equation}\label{Hidentity}
(fg^H + f^Hg)^H  = f^Hg^H - f g + f_0g_0\,,
\end{equation}
where $f_0$ and $g_0$ are subtraction constants of their corresponding functions, analogous to $\rho_0$ in (\ref{subtraction}). Since $\partial_x$ and the Hilbert transform commute, 
we find the Hilbert transform of (\ref{psieq}) as
\begin{equation}\label{psieqH}
(\lambda -1)\partial_x^2 \psi^H -2\pi\lambda\left[\rho_s^H\partial_x\psi^H - \rho_s \partial_x\psi + \rho_0\psi_0'\right] = 2m\omega\psi^H\,.
\end{equation}
 
 Life is made easier by the fact that $\psi'_0$, the subtraction constant in (\ref{psieqH}), needed to render the integral of $\psi'(x)$ over the real axis, is null. This should be expected, due to Cauchy's theorem, under the assumption that $\psi(x)$ is analytic in at least one of the half-planes where it also decays at infinity. We can also prove that 
$\psi'_0=0$ more directly. 
Since both $\psi(x)$ and $\psi'(x)$ are quasi-periodic and obey similar relations like (\ref{floquet}), it follows that 
\begin{eqnarray}\label{psi0}\int\limits_{-\infty}^\infty\psi'(x) dx  &=& \left(\int\limits_0^T\psi'(x) dx\right)\sum_{n=-\infty}^\infty e^{inqT} = 2\pi\left(\psi(T) - \psi(0)\right)\delta_P(qT)
\nonumber\\{}\nonumber\\
&=& 2\pi\psi(0)(e^{iqT}-1)\delta_P(qT)\,,
\end{eqnarray}
where $\delta_P(x)$ is the periodic delta-function of period $2\pi$. The last expression obviously vanishes, since $\psi(0)$ is finite by assumption.\footnote{Otherwise, if $\psi(0)$ diverges, we can pick any other period of $\psi(x)$ at which endpoints $\psi$ is finite.} Thus, $\psi'_0 = 0$ as promised and it can be dropped from (\ref{psieqH}).
Reversing the argument, this independent proof that $\psi'_0=0$ implies that if $\psi'(x)$ is analytic in one of the complex half-planes, it must decay to zero at infinity there. 

The natural thing to do now is to combine (\ref{psieq}) and (\ref{psieqH}), in a manner similar to (\ref{Riccati}). Thus, adding and subtracting $i$ times (\ref{psieq}) from (\ref{psieqH}), we obtain the pair of equations
\begin{equation}\label{complexified1}
(\lambda -1)\partial_x^2 F_{\pm} -2\pi\lambda\Phi_{\pm}\partial_x F_{\pm}  = 2m\omega F_{\pm}\,,
\end{equation}
where
\begin{equation}\label{Fpm}
F_{\pm}(x)  = \psi^H (x) \pm i\psi (x)\,.
\end{equation}

Recall that the two functions $\Phi_\pm (x)$,  defined in (\ref{Phipm}),  are the boundary values 
\begin{equation}\label{boundary}
\Phi_\pm(x) = \Phi (x\pm i0)
\end{equation}
of the meromorphic function $\Phi(z)$ in (\ref{PhiSolFinal}), as one approaches the real axis. Thus, $\Phi_+(x)$ is analytic in the $x-$UHP, and $\Phi_-(x)$ is analytic 
in the $x-$LHP. These are the coefficient functions in (\ref{complexified1}). It thus follows from the theory of differential equations that $F_+(x)$ should be 
analytic in the $x-$UHP, and that  $F_-(x)$ should be analytic in the $x-$LHP.

The two equations (\ref{complexified1}) can be united into a single equation
$$
(\lambda -1)F''(z) -2\pi\lambda\Phi (z)F'(z)  = 2m\omega F(z)\,,
$$
or more compactly, 
\begin{equation}\label{complexified}
F''(z) -{k\over\rho_0}\Phi (z)F'(z)  = {2m\omega\over \lambda -1} F(z)\,,
\end{equation}
for a single meromorphic function $F(z)$, such that
\begin{equation}\label{FF}
F_\pm(x) = F(x\pm i0)\,.
\end{equation}

Our first step in solving (\ref{complexified}) is to remove the first-derivative term. Thus, following standard methods, we substitute 
\begin{equation}\label{Ff}
F(z) = \xi (z) f(z)
\end{equation}
where $f(z)$ is a new unknown function, and $\xi(z)$ is determined by demanding that upon substituting (\ref{Ff}) in (\ref{complexified}), the coefficient of $f'(z)$ will be null. Thus, we find 
\begin{equation}\label{xieq0}
{\xi'\over \xi} = {k\over 2\rho_0}\Phi\,.
\end{equation}
Using (\ref{PhiSolFinal}) we may write this equation as 
\begin{equation}\label{xieq1}
{\xi'\over \xi} = {ik\over 2} + ik \,{ e^{ikz -u\,\sign(\Im z)}\over  1 - e^{ikz -u\,\sign(\Im z)}  }\,.
\end{equation}
The solution in each of the complex half-planes is immediate and is independent of the solution in the other half-plane. Thus,
\begin{equation}\label{xi}
\xi = i C_+\,\theta(\Im z)\,{ e^{ikz\over 2}\over  1 - e^{ikz -u}  } - i C_-\,\theta(-\Im z)\,{ e^{ikz\over 2}\over  1 - e^{ikz +u}  }\,,
\end{equation}
with integration constants $C_\pm$. In each half-plane, $\xi(z)$ is of course analytic, with discontinuity along the real axis. Granted more information on the analytic properties of our solution $\xi(z)$ and $F(z)$ we shall be able to determine these integration constants.

We also need an expression for $\xi''/\xi$ which appears in the equation for $f(z)$. Taking the derivative of (\ref{xieq0}) and using (\ref{Riccati1}) we find
\begin{equation}\label{xieq2}
{\xi''\over \xi} = {k\over 2\rho_0}\left(\Phi' + {k\over 2\rho_0}\Phi^2\right) = \left({k\over 2\rho_0}\right)^2\left(2\Phi^2 + \rho_0^2\right) \,.
\end{equation}
With (\ref{xieq2}) at our disposal, we finally obtain the very simple equation 
\begin{equation}\label{feq}
f''(z) + \left({k\over 2}\right)^2 f(z) = {2m\omega\over\lambda -1}f(z)
\end{equation}
for $f(z)$. We readily find the solution as 
\begin{equation}\label{f}
f(z) = e^{ipz}
\end{equation}
with $p$ a {\em real} momentum-like parameter.\footnote{More precisely, the two independent solutions of (\ref{feq}) are $e^{\pm ipz}$. 
Since they are related by flipping the sign of $p$, it is enough to consider only (\ref{f}), since we allow $p$ to range over positive and negative values.} 
The resulting dispersion relation is,  of course, 
\begin{equation}\label{omega1}
\omega(p) = {\lambda -1\over 2m}\left[\left({k\over 2}\right)^2 - p^2\right]\,.
\end{equation}
Finally, combining (\ref{f}), (\ref{xi}) and (\ref{Ff}) we obtain the solution of (\ref{complexified}) as
\begin{equation}\label{F1}
F(z) =  i C_+(q)\,\theta(\Im z)\,{ e^{iqz}\over  1 - e^{ikz -u}  } -  i C_-(q)\,\theta(-\Im z)\,{ e^{iqz}\over  1 - e^{ikz +u}  }\,,
\end{equation}
where
\begin{equation}\label{quasi}
q = {k\over 2} + p
\end{equation}
is the quasi-momentum, as indeed, $F(z+T) = e^{iqT}F(z)$, in accordance with (\ref{floquet}).  
In terms of $q$, the dispersion relation (\ref{omega1}) may be written as 
\begin{equation}\label{omega2}
\omega(q) = {\lambda -1\over 2m}q(k-q)\,.
\end{equation}
Note that $\omega(q)$ is independent of the amplitude of $\rho_s(x)$, which is governed by the parameter $u$. It depends on $\rho_s$ only through $k$, or equivalently, 
through $\rho_0$.

We are not done yet, since we have to impose the condition that $F_\pm(x) = F(x\pm i0)$ be consistent with (\ref{Fpm}) and also with the fact that $\psi'_0=0$ in (\ref{psieqH}), meaning that $\psi'(x)$ has to decay at infinity in the half-plane where it is analytic.  Bearing in mind the last requirement, we obtain from (\ref{Fpm}) that 
\begin{eqnarray}\label{psipsiH1}
\psi(x) &=& {F(x+i0) - F(x-i0)\over 2i}  = {C_+(q)\theta(q)\over 2}{ e^{iqx}\over 1 - e^{ikx -u}} + {C_-(q)\theta(k-q)\over 2} { e^{iqx}\over 1 - e^{ikx +u}}\nonumber\\{}\nonumber\\
\psi^H(x) &=& {F(x+i0) + F(x-i0)\over 2}  = {iC_+(q)\theta(q)\over 2}{ e^{iqx}\over 1 - e^{ikx -u}} - {iC_-(q)\theta(k-q)\over 2} { e^{iqx}\over 1 - e^{ikx +u}}\,.\nonumber\\{}
\end{eqnarray}
It would be perhaps useful to spend a few words to clarify these equations.  Let us concentrate on $\psi(x)$. The first term in $\psi(x)$, analytic in the $x-$UHP, decays there at infinity only for  $q>0$, which is guaranteed by the $\theta(q)$ prefactor. Similarly, the second therm in $\psi(x)$, analytic in the $x-$LHP, decays there at infinity only for $q<k$, which is guaranteed by the other prefactor $\theta(k-q)$. With this assignment of domains of quasi-momentum, the first term in $\psi(x)$ is an eigenvector of the Hilbert transform with eigenvalue $i$, and the second term is an eigenvector with eigenvalue $-i$, and therefore, $\psi^H(x)$ in the second line of (\ref{psipsiH1}), is indeed the Hilbert transform 
of $\psi (x)$. 

The domains of quasi-momentum in (\ref{psipsiH1}) appear asymmetric. In order to render them more symmetric, we shift $q$  in the second terms in both lines 
of (\ref{psipsiH1}) by one unit of the reciprocal lattice,  $q=q' + k$, where, of course, $q'<0$. After a little algebra (including absorbing a factor $-e^{-u}$ in $C_-$ and an appropriate redefinition thereof), and reinstating $q$ for $q'$, we can write $\psi(x)$ more symmetrically as
\beq\label{psi-final}
\psi(x)  =  {C_+(q)\theta(q)\over 2}{ e^{iqx}\over 1 - e^{ikx -u}} + {C_-(q)\theta(-q)\over 2} { e^{iqx}\over 1 - e^{-ikx -u}}\,.
\eeq
With this new assignment of quasi-momenta, the dispersion relation (\ref{omega2}) may be written more symmetrically as 
\begin{equation}\label{omega-final}
\omega(q) = {\lambda -1\over 2m}(k|q|-q^2)\,,
\end{equation}
where $q>0$ corresponds to the first term in (\ref{psi-final}), and $q<0$ to the second term. This function is made of two parabolas soldered together at $q=0$. 
It vanishes at $q=0, \pm k$, with a cusp at $q=0$. In addition, since ${\lambda -1\over 2m} >0$, it has two degenerate maxima at $q=\pm{k\over 2}$, with maximal value $\omega_{{\rm max}} = {\lambda -1\over 4m} k^2$. Therefore, each value of $\omega(q)$ between $\omega =0$ and $\omega_{\rm max}$ is four-fold degenerate, and occurs 
at $0<q<k$ as well as at $\pm (k-q)$ and $-q$ . All other possible values of $\omega(\pm q)$, with $|q|>k$,  are just doubly-degenerate. 

We readily identify $\om(q)$ as the dispersion relation of quasi-particles in the presence of {\em uniform} condensate $\rho_0$ \cite{ABJ:1995, BFM:2009}, as we have 
already mentioned following (\ref{zpe-BPS}).  In particular, it is independent of the parameter $u$, which governs the amplitude of oscillations of the BPS-soliton.

It is interesting to study the behavior of (\ref{omega-final}) in the limit $\lambda\rightarrow 1$. In this limit, according to (\ref{k}), $k$ diverges such that $(\lambda-1)k\rightarrow 2\pi\rho_0$, and therefore $\omega(q)\rightarrow 
{\pi\rho_0\over m}|q|$. This is, of course, the dispersion relation of sound waves propagating in the ground state of  a one-dimensional Fermi gas of uniform density
$\rho_0$ and Fermi momentum $k_F = \pi\rho_0$. This uniform Fermi gas configuration is the {\em only} solution of (\ref{BPS}) at $\lambda=1$.  
Thus, the spectrum (\ref{omega-final}) of fluctuations around the BPS-soliton (\ref{rhorhoH}) behaves smoothly as $\lambda$ goes through $\lambda=1$. This supports our assertion, made at the end of Section 2, that these BPS-solitons should also be taken to be smooth around $\lambda=1$, which is achieved by setting $e^u \gg {1\over \lambda-1}$ as $\lambda\rightarrow 1$.

 \subsection{The complete orthogonal set of Floquet-Bloch eigenfunctions}
Equipped with the explicit solution (\ref{psi-final}) (and the definition (\ref{psidef})), we can read-off the Floquet-Bloch wave-functions as
\beqra\label{Floquet-Bloch}
\phi_q(x) &=& \sqrt{{1-e^{-2u}\over 2\pi}\,{\rho_0\over\rho_s(x)}} { e^{iqx}\over 1 - e^{ikx -u}} \,,\quad q>0\nonumber\\{}\nonumber\\
\tilde\phi_q(x) &=& \sqrt{{1-e^{-2u}\over 2\pi}\,{\rho_0\over\rho_s(x)}} { e^{iqx}\over 1 - e^{-ikx -u}} \,,\quad q<0\,,
\eeqra
such that 
\beq\label{phiphitilde}
\tilde\phi_q = \phi_{-q}^*\,.
\eeq
Here, the normalization factors $C_{\pm}$ in (\ref{psi-final}) are chosen such that $\phi_q$ and $\tilde\phi_q$ be normalized to a delta-function, namely,
\beqra\label{normaization1}
\int\limits_{-\infty}^\infty\, dx\, \phi_q(x)\phi^*_p(x) & =& \delta (q-p)\quad q,p>0\nonumber\\{}\nonumber\\
\int\limits_{-\infty}^\infty\, dx\, \tilde\phi_q(x)\tilde\phi^*_p(x) & =& \delta (q-p)\quad q,p<0\,,
\eeqra
which can be easily established by using the factorization 
\begin{equation}\label{invrho}
{1\over \rho_s(x)} = {e^u\over 2\rho_0\sinh u}(1 - e^{ikx -u})(1 - e^{-ikx -u})\,.
\end{equation}
Our work is not done yet, since $\phi_q$ and $\tilde\phi_q$ do not form orthogonal sets.  To see this choose some $p>0$ and $q<0$ and consider
\begin{equation}\label{prod3}
\phi_p(x)\tilde\phi^*_q(x)  = {e^{-i(p+ |q|)}\over 2\pi} {1 - e^{-ikx -u}\over 1 - e^{ikx -u}} \,,
\end{equation}
where we used (\ref{invrho}) once more. The denominator can be expanded into a geometric series which we integrate term by term. Thus, we obtain
 \beqra\label{ortho3}
\int\limits_{-\infty}^\infty\, dx\, \phi_p(x)\tilde\phi^*_q(x)  &=&  \sum_{n=0}^\infty e^{-nu}
\left[\delta (p + |q| + nk) -e^{-u} \delta \left((n-1) k  + p  + |q|\right)\right]\nonumber\\{}\nonumber\\
 &=& - e^{-u} \delta (p  + |q| -k)\,,
\eeqra    
where in the last step we used the fact that $k, p$ and $|q|$ are all positive. In this region of parameters, the solutions of $p+|q|-k=0$ are any $0<p<k$ and $q=p-k$ (and therefore $-k<q<0$). Thus, $\phi_p(x)$, with $0<p<k$, and $\tilde\phi_{p-k}(x)$ are not orthogonal, and we need to rotate them into mutually orthogonal combinations. This rotation will mix the two states, which is allowed physically, since both states are degenerate in energy according to (\ref{omega-final}), and moreover, have the same quasi-momentum mod-$k$, i.e., both sates acquire phase $e^{ipT}$ upon spatial shift by one period $T$. Obtaining these orthogonal combinations is a straightforward task, which essentially amounts to diagonalizing Pauli's matrix $\sigma_x$.  The desired orthogonal combinations are found as
\beqra\label{phipm}
\phi_q^{(+)}(x) &=& {\phi_q(x) + \tilde\phi_{q-k}(x)\over \sqrt{2(1-e^{-u})}} = \sqrt{\rho_s(x)\over 4\pi\rho_0 (1 + e^{-u})}\,\left(e^{iqx} + e^{i(q-k)x}\right)\nonumber\\{}\nonumber\\
\phi_q^{(-)}(x) &=& {\phi_q(x) - \tilde\phi_{q-k}(x)\over \sqrt{2(1+e^{-u})}} = \sqrt{\rho_s(x)\over 4\pi \rho_0(1 - e^{-u})}\,\left(e^{iqx} - e^{i(q-k)x}\right)\,,\quad 0<q<k\,.\nonumber\\
\eeqra
It is easy to check, using (\ref{ortho3}), that they are normalized to a delta-function 
\beq\label{normaization1}
\int\limits_{-\infty}^\infty\, dx\, \phi^{(+)}_q(x)\phi^{(+)*}_p(x)  = \int\limits_{-\infty}^\infty\, dx\, \phi^{(-)}_q(x)\phi^{(-)*}_p(x)  = \delta (q-p)\quad 0<q,p<k\,.
\eeq
By construction, they are also orthogonal to all the other eigenfunctions, with quasi-momentum larger than $k$ or smaller than $-k$. 
In order to make our notations more symmetric with respect to assignment of quasi-momenta, let us rename in (\ref{phipm})
\beq\label{phiminus}
\tilde\phi_q^{(-)}(x) = - \phi_{k+q}^{(-)}(x)  = {\tilde\phi_q(x) - \phi_{q+k}(x)\over \sqrt{2(1+e^{-u})}} = \sqrt{\rho_s(x)\over 4\pi \rho_0(1 - e^{-u})}\,\left(e^{iqx} - e^{i(q+k)x}\right)\,,\quad -k<q<0\,,
\eeq
which of course, has no effect on the orthogonality relations. This concludes our derivation of the normalized orthogonal set of orthogonal Floquet-Bloch functions. For convenience, let us summarize them in the following list:
\beqra\label{Floquet-Bloch-Final}
\varphi_q(x) = \left\{
\begin{array}{cc}
\!\!\!\!\!\!\!\!\!\!\!\!\!\!\!\!\!\!\!\!\!\!\!\!\!\!\!\!\!\!\!\!\!\!\!\!\!\!\!\!\!\!\!\!\!\!\!\!\!\!\!\!\!\!\!\!\!\
\phi_q(x) = \sqrt{{1-e^{-2u}\over 2\pi}\,{\rho_0\over\rho_s(x)}} { e^{iqx}\over 1 - e^{ikx -u}} \,,
 &q>k\\{}\\
\phi_q^{(+)}(x) = {\phi_q(x) + \tilde\phi_{q-k}(x)\over \sqrt{2(1-e^{-u})}} = \sqrt{\rho_s(x)\over 4\pi\rho_0 (1 + e^{-u})}\,\left(e^{iqx} + e^{i(q-k)x}\right)\,,&0<q<k\\{}\\
\tilde\phi_q^{(-)}(x) =  {\tilde\phi_q(x) - \phi_{q+k}(x)\over \sqrt{2(1+e^{-u})}} = \sqrt{\rho_s(x)\over 4\pi \rho_0(1 - e^{-u})}\,\left(e^{iqx} - e^{i(q+k)x}\right)\,, &-k<q<0\\{}\\
\!\!\!\!\!\!\!\!\!\!\!\!\!\!\!\!\!\!\!\!\!\!\!\!\!\!\!\!\!\!\!\!\!\!\!\!\!\!\!\!\!\!\!\!\!\!\!\!\!\!\!\!\!\!\
\tilde\phi_q(x) = \sqrt{{1-e^{-2u}\over 2\pi}\,{\rho_0\over\rho_s(x)}} { e^{iqx}\over 1 - e^{-ikx -u}} \,,&q<-k\,,
\end{array}\right.\nonumber\\{}
\eeqra
for which 
\beq\label{noramlization-FinalA}
\int\limits_{-\infty}^\infty\, dx\, \vphi_q(x)\vphi^{*}_p(x)   = \delta (q-p)
\eeq
for all $q,p\in I\!\!R$, in accordance with (\ref{orthogonality}). 
\subsubsection{The completeness relation}
We shall now verify that the functions in (\ref{Floquet-Bloch-Final}) comprise a complete set. Thus, consider the LHS of (\ref{completeness2}), namely, 
\beq\label{Gamma}
\Gamma(x,y) =  \int\limits_{-\infty}^\infty\,dq \,\vphi^*_{ q}(x)\,\vphi_{ q}(y) \,,
\eeq
and split it into contributions of the four sets of functions in  (\ref{Floquet-Bloch-Final}) 
\beqra\label{GammaA}
\Gamma_1(x,y) &=&  \int\limits_k^\infty\,dq \,\phi^*_{ q}(x)\,\phi_{ q}(y) \nonumber\\{}\nonumber\\
\Gamma_2(x,y) &=&  \int\limits_0^k\,dq \,\phi_{q}^{(+)*}(x)\,\phi_{ q}^{(+)}(y) \nonumber\\{}\nonumber\\
\Gamma_3(x,y) &=&  \int\limits_{-k}^0\,dq \,\tilde\phi^{(-)*}_{q}(x)\,\tilde\phi_{ q}^{(-)}(y) \nonumber\\{}\nonumber\\
\Gamma_4(x,y) &=&  \int\limits_{-\infty}^{-k}\,dq \,\tilde\phi^*_{ q}(x)\,\tilde\phi_{ q}(y) \,.
\eeqra
From the first and fourth lines in (\ref{Floquet-Bloch-Final})  we obtain, by using the identities
\begin{eqnarray}\label{fourier}
\int\limits_k^\infty\,e^{iqx}\,dq &=& {ie^{ikx}\over x+i\epsilon}  = ie^{ikx}{P\over x} + \pi \delta (x)\nonumber\\{}\nonumber\\
\int\limits_{-\infty}^k\,e^{iqx}\,dq &=& {-ie^{ikx}\over x-i\epsilon}  = -ie^{ikx}{P\over x} + \pi \delta (x)\,
\end{eqnarray}
that 
\beqra\label{Gamma14}
\Gamma_1(x,y) &=& {1-e^{-2u}\over 2\pi}\,{|\rho_0|\over\sqrt{ \rho_s(x)\rho_s(y)}}\,{1\over (1-e^{-ikx-u})(1-e^{iky-u})}\,{ie^{ik(y-x)}\over y-x + i\epsilon}\nonumber\\
\Gamma_4(x,y) &=& \Gamma_1^*(x,y)\,,
\eeqra
where, the last line follows from (\ref{phiphitilde}). After some work we obtain
\beq\label{Gamma14A}
\Gamma_1(x,y) + \Gamma_4(x,y) = \delta(x-y) - {\sqrt{\rho_s(x)\rho_s(y)}\over \pi |\rho_0|(1-e^{-2u})}\,{\sin{k(x-y)\over 2}\over {x-y\over 2}}\,\left[\cos{k(x-y)\over 2} - e^{-u}\cos{k(x+y)\over 2}\right]\,,
\eeq
where we used the last equalities in (\ref{fourier}). In particular, the last cumbersome term in (\ref{Gamma14A}) arises from the principal parts in (\ref{Gamma14}), which combine in such a way that the singular principal part ${P\over x-y}$ is multiplied by a function which has a simple zero at $x=y$, which allows us to drop the $P$ symbol. 
$\Gamma_2(x,y)$ and $\Gamma_3(x,y)$ are evidently non-singular kernels. From the second and third lines  in (\ref{Floquet-Bloch-Final}), we can show,  after some straightforward but tedious calculation that $\Gamma_2(x,y) + \Gamma_3(x,y)$ exactly cancels the second, regular term in (\ref{Gamma14A}). Thus, 
\beq\label{completeness-Final}
\Gamma(x,y) = \delta(x-y)\,,
\eeq
proving completeness of the set of Floquet-Bloch functions (\ref{Floquet-Bloch-Final}). 

\subsection{Comments on uniqueness of the eigenstates}
Recall from the discussion following (\ref{omega-final}) that when $-k<q<k$,  $\om(q)$ is four-fold degenerate. Thus, e.g., for $0<q<k,$ also quasi-momenta 
$k-q, -q$ and $q-k$ all have a common value $\om(q)$ (which ranges between $\omega =0$ and $\omega_{\rm max}$ as $q$ varies in this domain). To this common 
eigenvalue correspond the four {\em orthogonal} eigenstates $\phi_q^{(+)}(x), \phi_{k-q}^{(+)}(x), \tilde\phi_{-q}^{(-)}(x) $ and $\tilde\phi_{q-k}^{(-)}(x) $. 
This quartet of states splits into two pairs of states with common quasi periodicity, namely, $\phi_q^{(+)}(x)$ and $\tilde\phi_{q-k}^{(-)}(x)$, which acquire phase $e^{iqT}$ 
under a shift of $x$ by one period $T$ (as can be seen from  (\ref{Floquet-Bloch-Final}), in accordance with (\ref{floquet})), 
and $\tilde\phi_{-q}^{(-)}(x) $  and $\phi_{k-q}^{(+)}(x)$, which acquire phase $e^{-iqT}$. We can rotate each pair of 
these states by a $2\times 2$ {\em unitary} matrix, which may even possibly be $q$-dependent, while leaving orthogonality (\ref{noramlization-FinalA}) and completeness 
(\ref{completeness2}) in tact. The rotated basis is as good as the original one given in (\ref{Floquet-Bloch-Final}). Thus, in this range of quasi-momenta, namely, $-k<q<k$, there is extra local {\em energy dependent} $SU(2)\times SU(2)$ symmetry. It would be interesting to investigate the origins of this symmetry further. 

In the remaining range of quasi-momenta, $|q|>k$, the energy spectrum is doubly degenerate, however, the corresponding two quasi-momenta $\pm q$ are not  generically separated by an integer multiple of $k$, and these states cannot be mixed, except for a discrete set of values $q_n = \pm {n\over 2} k$, with $n$ a positive integer, where there is an extra $SU(2)$ symmetry. 

Finally, let us briefly comment on the zero-energy solutions. Consider approaching $\om=0$ from within the domain of four-fold degeneracy. Let us pair the quartet of degenerate states according to their quasi-periodicities, as was discussed just above, namely, $(\phi_0^{(+)}(x),  \tilde\phi_{-k}^{(-)}(x))$ on one hand, and $(\tilde\phi_{0}^{(-)}(x), \phi_{k}^{(+)}(x))$ on the other. These four limiting states are of course strictly periodic, and we can mix them by a unitary transformation. Thus, the energy dependent 
$SU(2)\times SU(2)$ is enhanced at $\om=0$ to the much larger symmetry $SU(4)$.  However, let us return to the paired states of the lower $SU(2)\times SU(2)$ symmetry.
As can be seen from the explicit expressions in (\ref{Floquet-Bloch-Final}), we can take linear combinations of the members of each pair which are proportional to 
$\sqrt{\rho_s(x)\over\rho_0}$. For such combinations, according to (\ref{psidef}), we have $\psi(x) \propto \rho_s(x)$.  
It is straightforward to check directly that $\psi(x) = \rho_s(x)$ is indeed a solution of (\ref{ev2}) when $\omega = 0$. Indeed, upon substituting these $\psi$ and $\omega$ in (\ref{psieq}) we obtain the derivative of the BPS equation (\ref{BPS}) multiplied by $2\rho_s$. One physical origin of this zero-mode has to do with the translational collective coordinate of the BPS density wave - i.e., it can be shifted arbitrarily in space\footnote{Recall the imaginary parts $\Im u_+ = \Im u_-$ which we absorbed as a shift of $x$ following (\ref{PhiSol}).}. However, due to the enhanced $SU(4)$ symmetry, this cannot be the sole origin of zero eigenvalues in the spectrum of $H$, which requires further study.


\section{Conclusion and discussion}
\label{section5}
\setcounter{footnote}{1}

In this paper we have completely diagonalized the hamiltonian of quadratic collective field fluctuations in the background of BPS-density waves, which appear as 
static solutions of the equations of motion of the collective field formulation of the Calogero model. The fluctuation spectrum is positive, demonstrating linear stability of the 
BPS-density waves. Remarkably, the fluctuation spectrum $\om(q)$ around these BPS-density waves coincides with that of fluctuations around uniform condensates. 
The only difference between fluctuations around these two background types is the explicit form of the orthogonal complete set of Floquet-Bloch mode functions $\phi_q(x)$. 
We computed these functions explicitly for the BPS-density wave background, 
by diagonalizing explicitly a related periodic Schr\"odinger operator, with a non-standard term, containing the absolute value $|p|$ of the momentum operator.
Contrary to standard periodic Schr\"odinger operators, familiar from solid state physics, there are no gaps in the spectrum of the Schr\"odinger operator 
studied in this paper.

We close this paper by making some brief comments concerning fluctuations around non-BPS density wave solutions of the collective field equations
of motion. These variational equations, for static configurations $\rho(x)$, boil down to\cite{BFM:2009}
\beq\label{variationaleq} B[\rho]^2 - {\lambda
-1\over\rho}\,\partial_x (\rho B[\rho]) + 2\pi\lambda\,(\rho
B[\rho])^H - 2m\mu = 0\,,\eeq
where the BPS combination $B[\rho]$ was defined in (\ref{BPS}). Note that the BPS-density wave configuration $\rho_s(x)$, given in (\ref{rhorhoH}), is a solution 
of (\ref{variationaleq}) corresponding to vanishing chemical potential $\mu =0$. 
In \cite{BFM:2009} we have found solutions of (\ref{variationaleq}), which basically amount to subtracting from the BPS-density wave (\ref{rhorhoH}) either its minimum or maximum value, thus  obtaining extremal density waves which do vanish periodically, as opposed to the BPS-density wave. These solutions, which we named 
{\em vortex crystals} (corresponding to subtraction of the minimum, hence positive configurations ) and {\em anti-vortex crystals} (subtraction of the maximum, hence negative configurations), coincide with the large amplitude waves of \cite{Sen:1997qt} in the static limit.  We have discussed these vortex and anti-vortex crystals in \cite{BFM:2009} in detail, and computed the energy densities tied with them, which are all positive, hence above the zero-energy density tied with the BPS-density waves. 
Therefore the question of their linear stability poses an interesting problem, which is still open. 

This stability problem is more difficult to analyze than the stability of BPS-density waves, which we worked out here. To appreciate this difficulty, let us
derive the quadratic fluctuation hamiltonian in the background a non-BPS static solution $\rho$ of (\ref{variationaleq}). 
A convenient intermediate step is to expand 
\beq\label{Bfluct-x} B[\rho + \eta] = B[\rho] + \left[ {\lambda-1\over 2}\partial_x\left(\eta\over \rho\right) - \lambda\pi\eta^H\right] 
- {\lambda-1\over 2}\partial_x\left(\eta\over \rho\right)^2 +\ldots\,,\eeq
where the ellipsis stands for terms cubic in $\eta$ and higher.
Then, substituting (\ref{Bfluct-x}) in (\ref{Hcollective}), and expanding $H_{coll}$ 
to second order in $\eta(x)\,,\pi (x)$ and $\delta\mu$,  we obtain the quadratic piece $H_{coll}^{(2)} $ as 
\beqra\label{Hcoll1-x}
H_{coll}^{(2)}& =& V_{coll}[\rho]  -\delta\mu\,\int\,dx \,\eta + H_{sing}\nonumber\\{}\nonumber\\
&+&{1\over 2m}\,\int\,dx\, \rho(x)\,\left(\partial_x\pi(x)\right)^2 + 
{1\over 2m}\,\int\,dx\, \rho(x)\,\left[{\lambda-1\over 2}\partial_x\left(\eta\over \rho\right) - \lambda\pi\eta^H\right]^2\nonumber\\{}\nonumber\\
&-& {\pi\lambda\over m}\,\int\,dx \,B[\rho]\,\eta\eta^H \,,\eeqra
where we used the constraint (\ref{conservation}) and the variational equation (\ref{variationaleq}) to eliminate all terms linear in the shifted 
quantities $\eta$ and $\delta\mu$.  
Here $V_{coll}[\rho]$ is the energy tied in the extremal configuration $\rho(x)$ (which was null in the BPS case). 

Now we can see the crux of the problem: $\rho(x)$ is not a BPS configuration, and thus $B[\rho ] \neq 0$. This therefore generates a new term in (\ref{Hcoll1-x}), which did not arise in the BPS case, namely,  the last term in (\ref{Hcoll1-x}),  $-(\pi\lambda/m)\,\int\,dx \,B[\rho]\,\eta\eta^H\,.$ This term prevents us from carrying out the factorization 
of the quadratic pieces in (\ref{Hcoll1-x}) in a manner analogous to (\ref{Hcollbps}), which proved so useful to the complete and explicit diagonalization of the quadratic 
fluctuation hamiltonian around BPS-density waves. A new idea is clearly needed to solve the non-BPS fluctuation spectrum.

%

\appendix
\section{The collective field formulation of the Calogero model: derivation of $H_{coll}$ and its singular part $H_{sing}$  }
\label{app}
\setcounter{footnote}{1}
For the sake of being self-contained, we briefly review in this appendix the derivation of the collective field hamiltonian $H_{coll}$ in (\ref{Hcollective}), and its singular part 
$H_{sing}$ in (\ref{Hsing}).  Standard references on the collective field formalism are \cite{Jevicki:1979mb, sakbook}. Here we shall follow \cite{Feinberg:2004vz}, as well as 
\cite{Andric:1982jk, ABJ:1995},  which focus specifically on the Calogero model.

The Calogero model, whose quantum hamiltonian is given by 
\beq\label{calogero}
H = -{1\over 2m}\sum_{i=1}^N{\partial^2\over\partial x_i^2}+ {\lambda(\lambda-1)\over 2m}\sum_{i\neq j}{1\over (x_i-x_j)^2}
\eeq
describes $N$ identical particles of mass $m$ living in one dimension. These particles are subjected to inverse-square pair interactions with dimensionless
coupling $\lambda$. \footnote{Note that we did not include
in (\ref{calogero}) a confining potential. This is not really a
problem, as we can always add a very shallow confining potential
to regulate the problem (in the case of purely repulsive
interactions), or else, consider the particles confined to a very
large circle (i.e., consider (\ref{calogero}) as the large radius
limit of the Calogero-Sutherland model \cite{Sutherland:1971ep}).
We shall henceforth tacitly assume that the system is thus
properly regularized at large distances.}
The many-body wave functions are of the general form $\Psi(x_1,\ldots,x_N) = \Delta^\lambda S(x_1,\ldots, x_N)$, where $S(x_1,\ldots, x_N)$
is a function totally symmetric under any permutation of the particles, and $\Delta = \prod_{i<j}(x_i-x_j)$ is the Vandermonde determinant. 
The so-called Jastrow-factor $\Delta^\lambda$ arises due to the singular pair-interaction, which requires the wave function to vanish when any two particles 
coincide. The precise power-like vanishing is dictated by the requirement that the hamiltonian be self-adjoint.  

We see that for a generic value of $\lambda$, under 
the interchange of any two particles, the wave function suffers a phase change of $e^{i\pi\lambda}$. These particles are therefore anyons.  
The singular pair-wise interactions vanish, of course, at $\lambda=0$, which corresponds to non-interacting bosons,  and also at $\lambda=1$, which corresponds to non-interacting fermions. 

Powers of moments of the collective field $\rho(x)$ in (\ref{collective}) are clearly the building blocks of all functions which are totally symmetric in the particle coordinates. 
Thus, the symmetric factor in the  many-body wave-function, $S(x_1,\ldots, x_N)$, assuming it has a well-behaved large-$N$ limit, should  become a well-behaved functional 
of $\rho(x)$.  It is precisely these symmetric wave-functions $S(x_1,\ldots, x_N)$ on which the collective field operators, as well as the  collective hamiltonian $H_{coll}$ in (\ref{Hcollective}) act. Thus, in order to transform (\ref{calogero}) into a form amenable to collective field reformulation, we have to strip-off the Jastrow factors from the many-body wave function. 
This we achieve by performing on (\ref{calogero}) the similarity transformation
\begin{equation} \label{similarity}
H \rightarrow \tilde H = \Delta^{-\lambda} H \Delta^\lambda \,,
\end{equation}

It is straightforward to check that \cite{Andric:1982jk}
\beq\label{gauge-trick}
{1\over\Delta^\lambda} \left(\sum_{i=1}^N{\partial^2\over\partial x_i^2}  - \lambda(\lambda-1)\sum_{i\neq j}{1\over (x_i-x_j)^2}\right) \Delta^\lambda = 
{1\over\Delta^{2\lambda}}\sum_{i=1}^N{\partial\over\partial x_i} \Delta^{2\lambda} {\partial\over\partial x_i}\equiv \nabla_s^2\,.
\eeq
Thus, 
\beq\label{calogero1}
\tilde H = -{1\over 2m}\nabla_s^2\,.
\eeq
We can naturally interpret $\nabla_s^2$ as part of a laplacian in some set of curvilinear coordinates $q^a$, which in addition to the $x_i$,  also contain additional coordinates orthogonal to them. $\nabla_s^2$ is therefore the projection of the larger laplacian onto the subspace which depends exclusively on the original coordinates $x_i$. In particular, it is invariant under coordinate transformations which involve only the $x_i$.

The space parametrized by the coordinates $q^a$ is endowed with a metric $g_{ab}$. We need not concern ourselves with the details of this metric (and of the additional coordinates, orthogonal to the $x_i$) except for 
the following two facts:  First, we must clearly have\footnote{The volume element must be positive. Thus, we should really interpret $\Delta^{2\lambda} = (\Delta^2)^\lambda = \left|\Delta\right|^{2\lambda}$.}  
\beq\label{volelement}\sqrt{g} = \Delta^{2\lambda}\,.
\eeq 
Second, in the subspace of the original coordinates obviously $ds^2 = \sum_{i=1}^N dx_i^2$, and the corresponding block of $g_{ab}$ is therefore the unit matrix.

At the particular values $\lambda=\frac{1}{2}, 1$ and $2$, these curvilinear coordinates are those of the symmetric spaces corresponding to real-symmetric, complex-hermitian and quaternionic-self-dual matrices, respectively, and $\nabla_s^2$ is the projection of the laplacian into the singlet sector of the corresponding matrix space.

By construction, $\tilde H$ acts on the symmetric functions $S(x_i)$, and inner-products and matrix elements are computed with the integration measure
\beq\label{integrationmeasure}
d\mu = \Delta^{2\lambda}d^Nx\,,
\eeq
in respect to which $\tilde H$ is hermitian.

The collective field formalism amounts to performing a point canonical transformation from the $N$ position operators $x_i$ and their 
conjugate momenta $p_i = -i{\partial\over\partial x_i}$, to a new set of coordinates, namely, the collective field operators $\rho(x)$ and their conjugate 
momenta $\Pi(x) = -i{\delta\over\delta\rho(x)}$, and then expressing $\tilde H$ in terms of $\rho(x)$ and $\Pi(x)$.

\subsection{Point canonical transformations} More precisely, we should look upon this as 
a special case of point canonical coordinate transformations 
\beq\label{qQ}
q^a\rightarrow Q^a = Q^a({\bf q})
\eeq
of the larger space, parametrized by the entire collection of coordinates $q^a$, in which only the subspace parametrized by the $x_i$ 
is transformed into the new set of coordinates $\rho(x)$, while the subspace orthogonal to the $x_i$ remains unchanged. 

The metric in the new coordinates is given, in the usual manner, by 
\beq\label{newmetric}
\Omega_{ab}({\bf Q}) = g_{mn}({\bf q(Q)}){\partial q^m\over\partial Q^a}{\partial q^n\over\partial Q^b}
\eeq
and its inverse is given by 
\beq\label{invnewmetric}
\Omega^{ab}({\bf Q}) = g^{mn}({\bf q(Q)}){\partial Q^a\over\partial q^m}{\partial Q^b\over\partial q^n}\,.
\eeq
In these equations $q^a({\bf Q})$ are, of course,  the inverse coordinate transformations. 
It follows from (\ref{newmetric}) that
\beq\label{detOmega}
\Omega({\bf Q}) = \det \Omega_{ab}  = g J^2
\eeq
where
\beq\label{jacobian}
J = \det \left({\partial{\bf q}\over \partial {\bf Q}}\right)
\eeq
is the jacobian of the transformation. Thus, we have, 
\beq\label{newvolume}
\sqrt{\Omega} = \sqrt{g} J
\eeq
rendering the volume element invariant
\beq\label{invariantvolume}
\sqrt{\Omega} d{\bf Q} = \sqrt{g} d{\bf q}\,.
\eeq
Our next step is to transform the hamiltonian $\tilde H$ in (\ref{calogero1}) to the new coordinates. Since $\nabla_s^2$ in (\ref{gauge-trick}) is that part of the invariant laplacian 
in our space, which in particular, remains invariant under transformations that change only the $x_i$ while keeping the remaining orthogonal subspace unchanged, 
we simply have
\beq\label{Hnew}
\tilde H_Q = -{1\over 2m} \nabla_{Qs}^2\,,
\eeq
where 
\beq\label{nablaQ}
\nabla_Q^2 = {1\over\sqrt{\Omega}} {\partial\over\partial Q^a}\left(\Omega^{ab}\sqrt{\Omega}{\partial\over\partial Q^b}\right)\,,
\eeq
and the subscript $s$ in $\nabla_{Qs}^2$ means a projection on the subspace originally parametrized by the $x_i$. By construction, of course, $\nabla_{Qs}^2 = \nabla_s^2$.

This new expression for $\tilde H =\tilde H_Q$ is symmetric with respect to the measure $\sqrt{\omega}d{\bf Q}$. Life would be much easier if we could rid ourselves of this 
potentially complicated measure, and map $\tilde H$ onto an effective hamiltonian $H_{eff}$ which is symmetric with respect to the flat measure $d{\bf Q}$. This we can achieve by performing the  similarity transformation 
\beq\label{similarityH}
H_{eff} = \Omega^{1\over 4} \tilde H_{Qs}\Omega^{-{1\over 4}}\,.
\eeq

Let us now massage $H_{eff}$ into a more transparent form. In order to avoid cluttering of our formulas, we shall
henceforth not display the subscript $s$ explicitly. Thus, {\em all the formulas below should be properly projected onto the subspace originally parametrized by the $x_i$}.
It is a matter of straightforward calculation to show that 
\beqra\label{auxiliary}
\Om^{1\over 4}\,\nabla_Q^2\, \Om^{-{1\over 4}}  &=& 
\Om^{-{1\over 4}}\,{\pa\over\pa Q^a}\left(\Om^{ab}\sqrt{\Om}
{\pa\over\pa Q^b}\right)\, \Om^{-{1\over 4}}\nonumber\\
&=&\left(\Om^{-{1\over 4}}\,{\pa\over\pa Q^a} \Om^{1\over 4}\right)\Om^{ab}
\left(\Om^{1\over 4}{\pa\over\pa Q^b}\Om^{-{1\over 4}}\right)\nonumber\\
&=&{\pa\over\pa Q^a}\Om^{ab}{\pa\over\pa Q^b} - \left({\pa\over\pa Q^a}(
\Om^{ab}\,C_b)\right) - C_a\,\Om^{ab}C_b
\eeqra
where we have defined 
\beq\label{C}
C_a = {1\over 4} (\log \Om)_{,\,a}
\eeq
and where $(\cdot)_{,\,a}$ indicates a derivative with respect to $Q^a$. 
The operator $\Om^{1\over 4}\,\nabla_Q^2\, \Om^{-{1\over 4}}$ is manifestly 
symmetric with respect to the flat measure  $d{\bf Q}$, as is evident in each
of the lines in (\ref{auxiliary}). Thus, $H_{eff}$ is indeed the desired 
hamiltonian we set out to find, which, following (\ref{auxiliary}), we may 
write explicitly as
\beq\label{Hfinal}
H_{eff} = {1\over 2m} P_a\,\Om^{ab}\,P_b + {\hbar^2\over 2m} C_a\,\Om^{ab}\,C_b
 + {\hbar^2\over 2m}\left(\Om^{ab}\,C_b\right)_{,\,a}
\eeq
where we introduced the momentum operators 
\beq\label{momentumops}
P_a = -i\hbar {\pa\over\pa Q^a}\,,
\eeq
and displayed $\hbar$ dependence explicitly. The terms in (\ref{Hfinal}) quadratic in $\hbar $ may be thought of as
a generalization of the centrifugal barrier which arises in the radial 
hamiltonian in $D$ dimensions\footnote{The radial part of the $D$-dimensional 
laplacian $\nabla_r^2 = r^{-(D-1)}\pa_r (r^{D-1}\pa_r)$, defined with
respect to the measure $r^{D-1}dr$, may be transformed by a similarity 
transformation into $\tilde\nabla_r^2 = r^{D-1\over 2} \nabla_r^2 
r^{-{D-1\over 2}} = \pa_r^2 - {(D-1)(D-3)\over 4r^2} = 
\pa_r^2 - C_r\Om^{rr}C_r$, which is defined with
respect to the flat measure $dr$.}. Evidently, these terms are purely a 
quantum mechanical effect.

It is easy to see that
\beq\label{christoffel}
C_a = {1\over 2}\Gamma^b_{ba}
\eeq
where $\Gamma^a_{bc}$ is the second Christoffel symbol (i.e., the connection) 
of $\Om_{ab}$. However, sometimes a direct computation of the $C_a$ from their
definition (\ref{C}), or from the identity (\ref{christoffel}), may be too 
difficult to carry in practice. Thus, in order to bypass these potential 
difficulties, we shall now derive an identity satisfied by the $C_a$, from 
which we could compute them with somewhat less effort.

To this end we argue as follows: The invariant laplacian acting on a 
function which is a scalar under coordinate transformation produces yet 
another scalar function. Thus, 
\beq\label{laplaceianinvariance}
\nabla_q^2 \psi(q) = \nabla_Q^2 \psi(q(Q))\,.
\eeq
In particular, the coordinate functions themselves are scalars (their
differentials are one-forms). Let us define the quantities
\beq\label{omega}
\om^a =  - \hbar\nabla_q^2 Q^a = -\hbar\nabla_Q^2 Q^a\,.
\eeq
It follows from the definitions (\ref{omega}) and (\ref{C}) that 
$$\om^a = -\hbar\nabla_Q^2 Q^a = - {\hbar\over\sqrt{\Om}}
{\pa\over\pa Q^b}\left(\Om^{ab}\sqrt{\Om}\right) = -{\hbar\over 2}\Om^{ab}\,
(\log \Om)_{,\,b} - \hbar\Om^{ab}_{~,\,b}$$
or 
\beq\label{laplacianQ}
\om^a + 2\hbar\,\Om^{ab}C_b + \hbar\,\Om^{ab}_{~,\,b} = 0\,,
\eeq
which is the desired identity to determine the $C_a$. To simplify the 
computation, we are free to choose in (\ref{omega})
the coordinates $q^a$ in which the computation of 
$\om^a =  - \hbar\nabla_q^2 Q^a$ is as simple as possible. In our case, these are just the original coordinates.

\subsection{Application of the point canonical transformations to the collective field problem} 
The transformation from the $x_i$'s  to the density field $\rho(x)$ makes sense only in the limit
$N\rightarrow\infty$, since on one side there are $N$ position operators $x_i$, and a continuum of density operators $\rho(x)$ on the other side. The continuum of these density operators are not all independent. For example, they are subjected to the constraints (\ref{positiverho}) and (\ref{conservation}).    
Thus, in order to facilitate this transformation, one has to regularize the continuum theory. This is most conveniently achieved in momentum space. 
As our independent collective variables we choose the first $N$ Fourier modes 
\beq\label{rhok}
\rho_k = \int\,dx e^{-ikx}\rho(x) = \sum_{i=1}^N e^{-ikx_i}
\eeq
cut-off at some $k_{max}$, where $k$ is properly discretized, e.g., by putting our system in a large box of size $L$, and imposing appropriate boundary conditions. 
The details of this discretization are not important for our discussion of the large-$N$ limit. 
Assume next that the particles condense and that the mean particle density in the condensate is of the order of some value $\rho$. 
Thus, the microscopic inter-particle distance will be of the order $l\sim{1\over\rho}$. Consequently, the maximal Fourier
components should be of the order $k_{max} = {1\over l} \sim \rho$. Thus, the high density limit makes $k_{max}\rightarrow\infty$ (and letting $L\rightarrow\infty$ in the end,
makes $k$ continuous). The collective field reformulation of the model is therefore valid in the high density regime, where the system behaves like a {\em continuous  medium}. 

For large but finite spatial box size $L$, momenta are discrete. We then take the first lowest $N$ Fourier modes $\rho_k$ as our new coordinates $Q^a$. \footnote{As was stressed above, this transformation affects only the subspace originally parametrized by the $x_i$. The coordinates of the orthogonal complement subspace remain unchanged.} 
Since the metric $g_{ab}$ in the original coordinates is given by $ds^2 = \sum_{i=1}^N (dx^i)^2 +\ldots $, i.e., its block in the subspace of interest is simply the unit matrix, 
we obtain from (\ref{newmetric}) that 
\beq\label{newOmega}
\Omega^{\rho_k, \rho_{k'}} = \Omega(k,k'; [\rho]) = \sum_{i=1}^N {\partial \rho_k\over x_i}{\partial \rho_{k'}\over x_i}  = -kk'\rho_{k+k'}
\eeq
in obvious notations, where momentum modes play the role of the new coordinate indices.

The effective hamiltonian (\ref{Hfinal}) contains also the quantities $C_a$, 
which we will determine from the identity (\ref{laplacianQ}). Thus, we have to 
compute $\om^{\rho_k} = \om(k;[\rho])$. From the definition (\ref{omega})  
we obtain\footnote{From this point on we set
$\hbar=1$ again.} 
\beq\label{omgak1}
\om(k;[\phi])  = -\nabla_s^2\,\phi_k\,,
\eeq
where $\nabla_s^2$ was defined in (\ref{gauge-trick}). Thus, we obtain 
(see  Eqs.(3.11) -(3.13) in  \cite{Andric:1982jk})
\beq\label{omegak2}
\om(k;[\phi]) = 
k^2\sum_i e^{ikx_i} -2ik\lambda\sum_i e^{ikx_i}\sum_{j, j\neq i} {1\over x_i-x_j}\,.
\eeq
Adding and subtracting a $\sum_i$ term in the second term in (\ref{omegak2}) leads us to the final expression
\beq\label{omegak3}
\om(k;[\phi])  = (1-\lambda) k^2 \rho_k + \lambda k^2 \int \limits_0^1\,d\alpha \rho_{k\alpha}\rho_{k(1-\alpha)} \,.
\eeq

We should now substitute (\ref{newOmega}) and (\ref{omegak3}) in (\ref{laplacianQ}). It is easy to see that the last term there vanishes: 
\beq\label{vanishing}
\Omega^{ab}_{~,b} = -\sum_{k'} kk'{\delta\rho_{k+k'}\over \delta\rho_{k'}} =-\sum_{k'} kk'\delta_{k,0} = 0\,.
\eeq
Thus, we obtain from (\ref{laplacianQ}) that 
\beq\label{c1}
\omega[k;[\rho]) + 2\sum_{k'}\Omega(k,-k';[\rho]) C(k';[\rho]) = 0\,,
\eeq
which we can use to determine $C[k;[\rho])$. In the combined limits of large density and infinite spatial box, the $k$-sums tend to Fourier integrals. Thus, in the limit, 
\beq\label{c2}
\omega[k;[\rho]) + 2\int {dk'\over 2\pi}\,\Omega(k,-k';[\rho]) C(k';[\rho]) = 0\,,
\eeq
which we shall now transform to $x-$space. To this end we need (see Eq.(6) in \cite{Andric:1988vt} )
\beq\label{omegax}
\omega(x;[\rho])  = \int{dk\over 2\pi}\,e^{ikx} \omega(k;[\rho]) = (\lambda-1)\partial_x^2\rho(x) + 2\lambda\partial_x\pv\int\,{\rho(x)\rho(y)\over x-y}\,dy\,,
\eeq
as well as 
\beq\label{Omegaxy}
\Omega(x,y;[\rho]) = \int\,{dk dk'\over (2\pi)^2}\, e^{ikx + ik'y}\,\Omega(k,k';[\rho]) = \partial_x\partial_y \left(\rho(x)\delta(x-y)\right)\,.
\eeq
Using (\ref{omegax}) and (\ref{Omegaxy}), we Fourier transform (\ref{c2}) to $x$-space and obtain
\beqra\label{c3} 
&&\omega(x;[\rho]) -2\partial_x\left(\rho(x)\partial_xC(x;[\rho])\right) = \nonumber\\{}\nonumber\\
&&2\partial_x\left[{\lambda-1\over 2}\partial_x\rho(x) + \lambda\pv\int\,{\rho(x)\rho(y)\over x-y}\,dy  -\rho(x)\partial_xC(x;[\rho]) \right]=0\,,
\eeqra
which we can readily solve for $\partial_x C$, and obtain
\beq\label{cx}
\partial_x C(x;[\rho]) = {\lambda-1\over 2}{\partial_x\rho\over\rho} + \lambda\pv\int\,{\rho(y)\over x-y}\,dy\,.
\eeq

We now have all the ingredients required for computing the effective hamiltonian (\ref{Hfinal}). Thus, substituting (\ref{Omegaxy}), (\ref{cx}) and $P_a:= \Pi(x) = -i{\delta \over\delta\rho(x)}$ in the first two terms in (\ref{Hfinal}) we readily obtain the first two terms in (\ref{Hcollective}). The singular piece $H_{sing}$ in (\ref{Hsing}) arises from the last, divergence term  in (\ref{Hfinal}). From (\ref{laplacianQ}) and from (\ref{vanishing}) we thus have
\beq\label{Hsing1}
 {1\over 2m}\left(\Om^{ab}\,C_b\right)_{,\,a}  = -{1\over 4m} \om^a_{~,a}: =-{1\over 4m}\int\,dx {\delta\omega(x;[\rho])\over\delta\rho(x)}\,,
 \eeq
from which $H_{sing}$ in (\ref{Hsing}) follows.

\acknowledgments 

This work was supported in part by the Ministry of Science and Technology of the Republic of Croatia 
under contract No. 098-0000000-2865.  J.F. would like to thank the Rudjer Bo\v{s}kovi\'c Institute for its kind
hospitality, during which part of this work was completed.

\newcommand{\J}[4]{{\it #1} {\bf #2} (#3) #4}
\newcommand{\andJ}[3]{{\bf #1} (#2) #3}
\newcommand{\AP}{Ann.\ Phys.\ (N.Y.)}
\newcommand{\MPL}{Mod.\ Phys.\ Lett.}
\newcommand{\NP}{Nucl.\ Phys.}
\newcommand{\PL}{Phys.\ Lett.}
\newcommand{\PR}{ Phys.\ Rev.}
\newcommand{\PRL}{Phys.\ Rev.\ Lett.}
\newcommand{\PTP}{Prog.\ Theor.\ Phys.}
\newcommand{\hep}[1]{{\tt hep-th/{#1}}}

\end{document}